# Mathematical simulations of pediatric hemodynamics in isolated ventricular septal defect

Mitchel J. Colebank[1,2*], Alfonso Limon[3,4], Anthony Chang[3,4], Brandon Wong[4,5], Wyman Lai[4,5], Hamilton Baker[6]

[1]Department of Mathematics, University of South Carolina, Columbia, SC

[2]Department of Biomedical Engineering, University of South Carolina, Columbia, SC

[3]Mi4, Children's Hospital of Orange County, Anaheim, CA

[4]Children's Hospital of Orange County, Anaheim, CA,

[5]University of California, Irvine, Irvine

[6]Medical University of South Carolina, Charleston, SC

[*]Corresponding Author: mjcolebank@sc.edu

## Abstract

Computer modeling of the cardiovascular system has potential to revolutionize personalized medical care. This is especially promising for congenital heart defects, such as ventricular septal defect (VSD), a hole between the two ventricles of the heart. However, relatively few studies have built computer models for VSD, nor have they considered how natural adaptation to the cardiovascular system with age might interact with the presence of a small, medium, or large size VSD. Here, we combine a lumped parameter model of the cardiovascular system with two key modeling components: a size-dependent resistance dictating shunt flow between the two ventricles and age-dependent scaling relationships for the systemic and pulmonary circulations. Our results provide insight into changes in hemodynamic conditions with various VSD sizes. We investigate the combined effects of VSD size, vascular parameters, and age, showing distinct differences with these three factors. This study lays the necessary foundation for studying VSD and towards building digital shadows and digital twins for managing VSD in pediatrics.

## 1. Introduction

Mathematical models are an increasingly recognized future tool for the management of cardiovascular disease. The most notable example includes the use of computational fluid dynamics models to predict coronary artery hemodynamics and assist in subject-specific treatment of ischemic conditions[1]. These computational frameworks have potential in managing complex congenital heart diseases[2–5]. While fluid dynamics models have high spatial fidelity, they often come at elevated computation time which may be infeasible for clinical decision making. Thus, lumped-parameter models, which describe flow through compartments and interactions between cardiac and vascular segments, are attractive alternatives. These models have also received attention in predicting neonatal and pediatric hemodynamics under the effects of congenital heart defects and surgical correction[6–10].

The most common congenital heart defect is ventricular septal defect (VSD), which affects nearly 3% of all live births and accounts for roughly 1/3 of all congenital heart disease diagnoses[11–13]. The defect, a hole in the interventricular septum, which separates the left ventricle (LV) and right ventricle (RV), can drastically change cardiac and vascular hemodynamics depending on the defect size and location.  Small, high-resistance VSDs (typically < 3 mm in diameter at birth) often close spontaneously with increasing age of the subject, whereas subjects with large VSDs at birth (> 6 mm in diameter) undergo surgical intervention[11], as they can cause congestive heart failure, impaired growth, and eventual pulmonary vascular disease if left untreated. Medium-sized VSDs (3 to 6 mm in diameter at birth) may or may not close naturally depending on the hemodynamic status of the subject. The latter can cause growth and remodeling and cardiovascular complications (e.g., pulmonary hypertension[11,14,15]). Gold standard diagnosis and management of VSD uses echocardiography measurements to assess VSD size, pulmonary-to-systemic flow (Qp/Qs) ratio, left atrial and LV size, and pulmonary hemodynamic metrics to assess appropriate standard of care[11,14–16]. However, VSD diagnosis is associated with significant increases in mortality due to heart conditions at late age points[16]. To date, relatively few computational studies have simulated hemodynamics in the setting of a VSD[9,10], nor have studies considered how natural adaptation with growth in combination with a VSD alters hemodynamics.

To address this knowledge gap, this study uses a lumped-parameter model of the cardiovascular system with age-based parameter adaptation rules to simulate the effects of VSD. We combine an eight-compartment model with previously established allometric- and maturation-based parameter scaling of the systemic and pulmonary vasculature[8]. We simulate hemodynamics from one day to two years of age in a normal physiological pediatric case, as well as in small, medium, and large VSDs. We consider the effects of elevated pulmonary vascular resistance (PVR), consistent with the development of pulmonary hypertension, as well as uncertainty in maturation-based adaptation and total blood volume. Our results provide an initial step towards simulating age-dependent hemodynamic function with isolated VSD, a necessary step towards using predictive modeling in managing congenital heart diseases.

## 2. Materials and Methods

### 2.1 Mathematical model of the cardiovascular system and VSD

This study utilizes systems of ordinary differential equations (ODEs) that simulate dynamic pressure $p(t)$ (mmHg), volume $V(t)$ (mL), and flow $q(t)$ (mL/s) using reformulations electrical circuit equations[17,18]. The model includes four heart chambers: the left atrium (LA), left ventricle (LV), right atrium (RA), and right ventricle (RV). We also include systemic arterial (SA), systemic venous (SV), pulmonary arterial (PA), and pulmonary venous (PV) compartments for the circulation. The two atrioventricular (tricuspid and mitral) and two semilunar (pulmonary and aortic) valves are modeled as diodes, closing when the pressure gradient across the valve is negative. A schematic of the model is provided in Figure 1. Though this neglects inertial effects of valve dynamics, it can recapitulate valvular-flow profiles in the absence of regurgitation. The differential equations for volume and algebraic relationships for compartment and valve flow, compartment pressure, and heart pressure are given by

$$\frac{dV_i}{dt} = q_{in} - q_{out}, \tag{1}$$

$$q_{c,i} = \frac{\Delta p_i}{R_i}, \quad q_{val_i} = \max\left(0, \frac{\Delta p_i}{R_i}\right) \tag{2}$$

$$p_{c,i} = \frac{V_{c,i}(t)}{C_i}, \tag{3}$$

$$p_{h,i}(t) = E_{h,i}(t)V_{h,i}(t) \qquad (4)$$

respectively. Equation (1) describes changes in stressed volume $V_i(t)$ (the blood volume that generates transmural pressure) in compartment $i$, which is dictated by flow conservation using equation (2). Non-valve flow dynamics are proportional to the ratio of the pressure gradient $\Delta p_i$ and the resistance, $R_i$ (mmHg s/mL) between compartments[18]. Valve flow equations are similar but set to zero when pressure gradients become negative[17]. Pressure dynamics are modeled differently in four vascular compartments, $p_{c,i}$, and four heart chambers, $p_{h,i}$. The pressure within a vascular compartment is the ratio of volume ($V_{c,i}$) and compliance, $C_i$ (mL/mmHg).

Pressures in each heart chamber, $p_{h,i}$, are the product of cardiac elastance, $E_{h,i}$ (mmHg/mL), and the blood volume in the heart, $V_{h,i}$. The elastance function driving heart contraction is described by a piecewise cosine function[17]

$$E_{h,i}(t) = \begin{cases} E_{min,i} + \dfrac{E_{max,i} - E_{min,i}}{2}\left(1 - \cos\left(\dfrac{\tilde{t}}{T_{max,i}}\right)\right), & \tilde{t} \in [0, T_{max,i}] \\ E_{min,i} + \dfrac{E_{max,i} - E_{min,i}}{2}\left(1 + \cos\left(\dfrac{\tilde{t} - T_{max,i}}{T_{min,i} - T_{max,i}}\right)\right), & \tilde{t} \in (T_{max,i}, T_{min,i}] \\ E_{min,i}, & else \end{cases} \qquad (5)$$

where $\tilde{t} = \text{mod}(t, T)$ is the cardiac cycle time modulated by the cardiac cycle length, $T = 60/HR$ (s), where $HR$ is the heartrate in beats per minute. The elastance function oscillates between the minimum elastance, $E_{min,i}$, and maximum elastance, $E_{max,i}$, for each heart compartment $i = LA, LV, RA, RV$. An additional offset parameter, $T_{off}$ (s), is prescribed to the atria so that atrial contraction precedes ventricular contraction[17].

To simulate flow through a VSD, we include an additional resistor between the LV and RV[19]. The flow across the ventricular defect, $q_{VSD}$, is modeled by reduced order orifice flow

$$q_{VSD} = \text{sign}(\Delta p) \cdot \pi r_{VSD}^2 \sqrt{p^* \cdot (2|\Delta p|/\rho)}, \qquad \Delta p = p_{LV}(t) - p_{RV}(t) \qquad (6)$$

where $r_{VSD}$ is the VSD radius (cm), $\rho = 1.055$ (g/cm$^3$) is the blood density, and $p^* = 1333.22$ (g/cm s$^2$) is a conversion factor between mmHg and centimeter-grams-seconds (CGS) units. The sign convention implies that flow from the LV to the RV is considered positive while flow from the RV to the LV is considered negative. The volume equations for the LV and RV are modified to

$$\frac{dV_{LV}}{dt} = q_{M,Val} - q_{A,Val} - q_{VSD}, \quad \text{and} \tag{7}$$

$$\frac{dV_{RV}}{dt} = q_{T,Val} - q_{P,Val} + q_{VSD}, \tag{8}$$

respectively, where $q_{M,Val}$, $q_{A,Val}$, $q_{T,Val}$, and $q_{P,Val}$ are the mitral, aortic, tricuspid, and pulmonic valve flows (mL/s), respectively. We parameterize the VSD resistance element as $R_{VSD} = \left(\pi r_{VSD}^2 \cdot \sqrt{p^* \cdot 2/\rho}\right)^{-1}$ by reformulating equation (6) and use this resistance in our analysis. All model parameters and outputs are computed in units of mmHg, mL, and seconds.

> Figure 1. (a) Schematic of the mathematical model. The left atrial (LA), left ventricle (LV), right atrium (RA), and right ventricle (RV) are described by elastance functions, and connected to the systemic and pulmonary arteries (SA, PA) and veins (SV, PV) through resistors. The circulation is modeled as a compliance chamber. (b-c) Growth trajectories for female and male individuals (respectively) through one year of life based on CDC data.

2.2 Parameterization of the model

The system of ODEs is driven by resistance, compliance, elastance, and timing parameters. We begin by parameterizing the model for a reference adult of either sex. We then scale the adult parameters to neonate and pediatric values, as discussed in Section 2.3. A table describing the parameters and their *a priori* calculations can be found in Table 1.

We assume that the adult blood volume is 4500 mL in females and 5000 mL in males[20]. Each model compartment is given a percentage of the total blood volume. A fraction of this is considered the stressed, circulating volume, similar to previous studies[17]. A list of total volumes used as initial conditions for the model can be found in Table 1. We assume that the pressures are similar between female/male, with slightly lower systemic and pulmonary pressures in females consistent with the literature[20]. We also assume a stroke volume of 70 mL in adult females versus 80 in adult males, and heart rates of 80 and 70 beats per minute for females and males, respectively. This provides a cardiac output of 5.6 L/min (or 93.33 mL/s) for females and males. We then use reference values for compartment pressures and literature derived distributions for the total blood volume[17,20] to construct initial parameter estimates for the model.

Resistance values, $R_i$, are set as the pressure gradient across the compartment divided by the cardiac output. Valvular resistances are fixed at smaller values to ensure proper filling during diastole and avoid pressure differentials seen in valvular stenoses[17]. The resistance of the VSD is set to $1 \times 10^{10}$ (mmHg/mL) in the absence of a defect or defined as shown in section 2.1. Systemic and pulmonary arterial compliances, $C_{sa}$ and $C_{pa}$, are defined as stroke volume over pulse pressure (i.e., the difference between systolic and diastolic pressures). Systemic and pulmonary venous compliances, $C_{sv}$ and $C_{pv}$, are defined as compartment volume divided by the mean pressure in the respective compartments. The ventricular end-diastolic elastance, $E_{min}$, and end-systolic elastance, $E_{max}$, are defined as $E_{min} = p_{dias}/V_{dias}$ and $E_{max} = p_{sys}/V_{sys}$, respectively. For atria, we use the same relationship for $E_{max}$, but instead use $E_{min} = (p_{sys} + p_{dias})/(V_{sys} + V_{dias})$ for minimum atrial elastance to provide larger pressure developments during atrial filling[17]. We assume that peak contraction, $T_{max}$, and the start of cardiac diastole, $T_{min}$, occur 25% into the cardiac cycle and 55% into the cardiac cycle, i.e. $T_{max} = 0.25 \cdot T$ and $T_{min} = 0.55 \cdot T$, respectively. We assume that atrial contraction occurs $0.15 \cdot T$ before ventricular contraction, effectively making $T_{max} = 0.95 \cdot T$ and $T_{min} = 0.05 \cdot T$ in the atria. Table 1 provides a summary of the model parameters, their units, interpretations, and calculation methods.

2.3 Age-dependent parameterization

We modify the framework presented by Hiebing et al.[8] for generating age specific parameterization using allometric and nonlinear scaling of the adult parameters. To begin, we fit a cubic function to pediatric age-weight data from the Centers for Disease Control online database to provide the 25th, 50th, and 75th percentile weight for any age group within the first two years of life. We include both male and female growth trajectories, though the general shape of the growth curves (shown in Figure 1) are nearly identical, with females have lower weight magnitudes. Allometric scaling is driven by these weights using

$$X_i = X_0 \cdot \left(\frac{M_i}{M_0}\right)^b \tag{9}$$

where $X_i$ is the scaled parameter, $X_0$ is the reference value, and $M_i$ and $M_0$ are the current and initial body weights (kg) of the patient. The exponent $b$ varies between different parameters of the model, as described in Table 2. We assume that $X_0$ and $M_0$ correspond to the adult parameterization, shown in Table 1. We employ this allometric scaling relationship for the total blood volume total blood volumes, and all model parameters except pulmonary and systemic arterial resistance and compliance.

The maturation approach presented by Hiebing et al.[8] simulates the nonlinear changes in pulmonary and systemic arterial parameters during the initial transition of the neonatal and pediatric circulation. The scaled PVR, $R_p$, and systemic vascular resistance (SVR), $R_s$, are described as a function of time $t$ in days

$$R_p(t) = \frac{g_{p,1}}{1 + (t/g_{p,2})^{g_{p,3}}} + R_p(0) \cdot \left(\frac{M_i}{M_0}\right)^{b_{R_p}} \cdot \left(\frac{(t/g_{p,2})^{g_{p,3}}}{1 + (t/g_{p,2})^{g_{p,3}}}\right) \quad (10)$$

$$R_s(t) = \frac{g_{s,1}}{1 + (t/g_{s,2})^{g_{s,3}}} + R_s(0) \cdot \left(\frac{M_i}{M_0}\right)^{b_{R_s}} \cdot \left(\frac{(t/g_{s,2})^{g_{s,3}}}{1 + (t/g_{s,2})^{g_{s,3}}}\right). \quad (11)$$

We use a different mathematical model than Hiebing et al.; hence, we modify the $g$ parameters from equations (10) - (11) to fit our approach. This provides $g_{p,1} = 4.0$ (mmHg s / mL), $g_{p,2} = 1.0$ (days), and $g_{p,3} = 2.0$ for the pulmonary circulation, and $g_{s,1} = 3.8$ (mmHg s / mL), $g_{s,2} = 50$ (days), and $g_{s,3} = 1.1$ for the systemic circulation. We also set $b_{pa} = -0.75$ like the other resistance parameters, but use $b_{sys} = -0.5$ for a more gradual change in SVR over time. These values provide hemodynamic predictions from the model that are consistent with the literature over two years of life (shown later)[21–29]. We use equations (9) - (11) to provide age-dependent parameters as a function of age for model simulations across two years of life.

## 2.4 Simulating VSD with pulmonary vascular dysfunction

One possible complication of VSD is elevated $R_p$ and reduced $C_{pa}$ with aging. This is attributed in part to elevated Qp/Qs ratios, where large pulmonary flow increases shear stress magnitude on the pulmonary arteries and causes out-of-proportion increases in resistance due to endothelial dysfunction[30,31]. We investigate this in two scenarios. In the first, we assume pulmonary vascular parameters adapt using equation (10) during the first 30 days of life, but that $R_p$ and $C_{pa}$ then remain at the fixed value, $R_p(30)$ and $C_{pa}(30)$, while other parameters adapt with age. In the second case, we assume that pulmonary adaptation only occurs up until three days of age, and fix $R_p$ and $C_{pa}$ at $R_p(3)$ and $C_{pa}(3)$, respectively. We investigate how these changes affect hemodynamics at time points up to two years of age.

2.5 Quantifying uncertainty in vascular adaptation

The maturation functions in equations (10) - (11) dictate how parameters adapt with age but are based on observations across multiple datasets. Thus, we investigate the predictive uncertainty in our model by sampling the maturation parameters, as well as the total blood volumes, and simulating growth trajectories. We simulate 100 different trajectories from birth to two years of age by sampling each $g$ parameter from equations (10) - (11) from a uniform distribution with bounds $\pm 30\%$ of the nominal value, i.e.

$$g_{i,j}^k \sim U\left(\frac{2g_{i,j}}{3}, \frac{4g_{i,j}}{3}\right), \qquad i = p, s, \qquad j = 1,2,3 \qquad (12)$$

where $g_{i,j}$ is the nominal value of the maturation parameters for the pulmonary or systemic system. We apply the same uncertainty bounds to the total blood volume to simulate differences in subject size. We examine how time-dependent cardiac function (i.e., pressure and volume curves) and scalar metrics (mean pulmonary artery pressure (mPAP), mean systemic arterial pressure (MAP), and the ratio of cycle averaged pulmonic and aortic flow, Qp/Qs) vary when adaptation curves are treated as uncertain across two years of age. We encode and simulate the mathematical model in MATLAB (MathWorks, Nantick MA) version 2024a using the *ode15s* solver with relative and absolute tolerances of $1 \times 10^{-6}$, using 30 heartbeats to achieve convergence.

## 3. Results

3.1 Comparison of the pediatric model to data and adult simulations

Figure 2 shows the predicted MAP and mPAP from the mathematical model over two years of life in both sexes with adaptation according to equations (9) - (11). We see that MAP increases with age while mPAP has a drastic drop during the first several days of development consistent with reduced PVR. These predictions are compared to literature data[21–29] as originally presented by Hiebing. Overall, model predictions are within the variability reported from previous studies. We also show end-systolic pressure (ESP), end-systolic volume (ESV), end-diastolic pressure (EDP), and end-diastolic volume (EDV) in both ventricles in Figure 2(c)-(f). Both heart chambers show an increase in ESV and EDV, with minor sex-differences. Predictions of LV ESV and RV ESV are smaller than data reported for subjects closer to two years of age.

Figure 2. Changes in (a) MAP and (b) mPAP over the first two years of life in both females and males. Data from multiple studies[21–29] reproduced in the study by Heiberg are also presented. A subplot of mPAP during the first two weeks of life is also provided in (b). End-diastolic volumes (EDV, (c) and (d)) and end-systolic volumes (ESV, (e) and (f)) are presented over the same time-frame as subplots (a) and (b). Data from multiple studies[21–29] reproduced in the study by Hiebing are also presented.

We compare our model results between a male adult and a male newborn (two days of age) in Figure 3. Adult simulations in Figure 3(a) show physiologically consistent blood pressure predictions and realistic, biphasic tricuspid and mitral valve flow patterns[20]. Atrial pressure-volume loops do not capture the more complex "a" and "v" loops seen in typical LA and RA dynamics. Pediatric simulations show several differences from adult hemodynamics. Cardiac volumes are significantly smaller, consistent with smaller heart size. Equations (10) - (11) enforce a lower SVR and higher PVR with earlier age, resulting in a decreased LV systolic pressure and elevated RV systolic pressure, shown in Figure 3(d). Pulmonary arterial pulse pressure in Figure 3(e) is higher, consistent with a small pulmonary vascular compliance. Flow simulations in Figure 3(f) are similar in shape to the adult simulations in Figure 3(c), but with substantially smaller magnitudes. Heart rate is also faster in the pediatric case.

Figure 3. (a-c) Simulations in a typical adult, 70 kg Male, including atrial and ventricular pressure-volume loops (a), vascular pressures (b), and valvular flows (c). (d-f) Simulations from the same model but with parameterizations based on allometric and maturation-based scaling laws to a two-day old. Note that pulmonary pressures are significantly higher, with smaller cardiac volumes. See Figure 1 for abbreviations for the model infrastructure.

3.2 Simulating VSD, growth, and the effects of pulmonary vascular dysfunction

We simulate a seven-day old female (Figure 4(a)-(b)) and male physiology (Figure 4(c)-(d)) across four cases: no VSD, and small, medium, or large VSD. The latter three are parameterized by VSD diameters of 1, 3, and 6 mm, respectively. Hemodynamics are similar between no VSD and a small VSD across both sexes. Medium-sized VSDs increase RV pressures and slightly decrease LV pressures, with a small rightward shift in RV volumes shown in Figure 4(a)-(b). There is also an increase in LA pressure and volume, whereas RA hemodynamics minimally change. Large-sized VSDs amplify prior hemodynamic changes, with systolic pressures in the two ventricles approaching a similar magnitude. A large VSD increases RV EDV, while LV stroke volume increases along with LV EDV. There is a slight sex difference in RV EDV, with larger values in males.

Valvular and VSD flow are shown in Figure 4(c)-(d). Left-to-right shunts (indicated by positive VSD flow) increase with larger VSD sizes. Though tricuspid flows are similar, mitral valve flow is larger in magnitude for larger VSDs. Large VSDs also increase pulmonic valve flow and reduce the time during which the aortic valve is open, leading to larger total pulmonary flow relative to systemic flow (i.e., Qp/Qs).

Figure 4. Simulations in a seven-day old pediatric female (a,c) and male (b,d) subject with and without a VSD. Cardiac function (a-b) and valvular flows (c-d) are shown for each sex with no VSD, a small VSD, a medium VSD, and a large VSD, corresponding to 0, 1, 3, and 6 mm in diameter, respectively. Subplots (c) and (d) include flow across the VSD, with positive flow indicated left-to-right shunting.

We investigated how VSD size and PVR may impact simulated cardiac and hemodynamic metrics. Results in Figure 5 show six simulated scalar metrics against different PVR magnitudes for a 30-day old Male subject with various VSD sizes. These include the Qp/Qs ratio, mPAP, and cardiac work in the LV, RV, LA, and RA. The PVR values are presented in Wood units (WU), which are equivalent to mmHg·min / L. Cardiac work was defined as the area within the pressure-volume loop and converted to Joules using the conversion factor 1 J = $7.5 \times 10^3$ mmHg·mL. The Qp/Qs magnitude is significantly higher in medium and large VSDs compared to small or no VSD. As PVR magnitude increases, we get a decrease in Qp/Qs such that at 30 WU, even large VSDs fall below the 1.5 ratio used clinically for VSD classification[11]. However, such values of PVR are quite high and represent a severe pathology. Larger PVR values increase mPAP, with pulmonary hypertensive conditions defined for mPAP > 20 (provided in Figure 5(b)). The difference in mPAP between small and large VSDs is apparent for low PVR, but larger PVR values diminish this difference.

Results in Figure 5(c)-(f) show cardiac work in all four heart chambers. LV cardiac work is smallest in the case of a large VSD, consistent with the largest left-to-right shunt. Increasing PVR reduces LV cardiac work. The opposite holds for the RV, which has significantly higher cardiac work for a large defect. Increasing PVR causes a decrease in LV work and an increase in RV work. Work by the LA is highest for low PVR and a large VSD. Large VSDs increase RA work as well, but RA work decreases with higher PVR.

Figure 5. Metrics of hemodynamic and cardiac function for various VSD size (provided as diameter) and PVR magnitudes, presented in Wood Units (WU), for a 30 day old male subject. (a) Ratio of mean pulmonary flow to mean systemic flow (Qp/Qs), with the threshold of 1.5 provided. (b) Mean pulmonary arterial pressure (mPAP), with the threshold of 20 mmHg used to define pulmonary hypertension. (c-f) LV, RV, LA, and RA cardiac work calculated from simulated pressure-volume loops.

We simulated aging using the allometric and maturation approaches in equations (9) – (11) across non-VSD and small-, medium-, and large-sized VSDs. Figure 6 shows LV and RV pressure-volume loop simulations in a female subject. Both LV pressures and volumes tend to increase with age. The LV exhibits increased EDVs and stroke volumes with larger defect sizes.

Systolic pressures in the LV at two years of age are noticeably lower with a large VSD than no VSD. In the RV, we see elevated pressures in all VSD conditions across the first three days of life. There is a clear increase in EDV and ESP in the RV with larger VSDs at three days of age. There is an abnormal curve to the RV PV loop during isovolumic contraction, which is attributed to increased volume through the left-to-right shunt. Predictions at day 30, 60, and 90 show more subtle differences between no VSD, small-, and medium-size VSD predictions. The RV ESP is clearly elevated in medium- and large-sized VSDs, even though $R_p$ is the same.

Figure 6. Ventricular pressure-volume loops as a function of age and VSD severity in a female. Parameters of the model are scaled using allometric and maturation-based approaches as described in the text. (a) Hemodynamics in the LV from one day to two years of age. VSD diameters used in the simulations are 0, 1, 3, and 6 mm in diameter. (b) Hemodynamics in the RV over the same time

We conducted a similar set of simulations in Figure 6 but with different pulmonary vascular adaptation scenarios. The first set of simulations (Figure 7(a)-(b)) use the maturation function to progress until day 30, at which time the value of $R_p(t) = R_p(30)$ and $C_{pa}(t) = C_{pa}(30)$. The second set of simulations (Figure 7 (c)-(d)) allow pulmonary adaptation until three days of age. In the former case (Figure 7(a)-(b)), we see relatively little change in the LV compared to results in Figure 6. In contrast, the RV undergoes a relatively large drop in pressure from one day to 30 days of age, but then experiences an increase in pressure magnitudes at later age points. This is more prominent for the large-sized VSD, where there is an accompanying increase in RV EDV with larger defects. For the second set of simulations, we see that LV EDV is similar with all VSD sizes, with only minor changes in pressure (Figure 7(c)). The RV predictions (Figure 7(d)) show an elevated ESP across all four time points, with a slight decrease in pressure for small-and medium-sized VSDs at 30 days of age. The RV EDV values for no-VSD and the three VSD sizes are similar when PVR remains high.

Figure 7. Ventricular pressure-volume loops as a function of age and VSD severity with different pulmonary vascular conditions in a female. All parameters of the model are scaled as described in the text except for $R_p$ and $C_{pa}$. (a-b) Hemodynamics in the LV and RV from 1 day to two years of age under the assumption that $R_p$ and $C_{pa}$ change until 30 days of age. (b) Corresponding RV predictions. (c-d) Hemodynamics in the LV and RV from 1 day to two years of age under the assumption that $R_p$ and $C_{pa}$ are fixed at the value obtained at 3 days of age.

3.3 Sampling based uncertainty propagation for age-dependent hemodynamics

Finally, we consider how uncertainties in the maturation parameters of equations (10) - (11) and total blood volume may change the trajectory of hemodynamics as a function of age. Arterial resistance and compliance trajectories are presented in Figure 8, which show larger uncertainty in $R_p, R_s,$ and $C_{pa}$ during the initial stages of the maturation curve. In contrast, $C_{sa}$ has relatively constant uncertainty throughout two years of life. We subsequently propagated these uncertainties through the model to provide individual trajectories up to two years of age for each curve presented in Figure 8. We did this in small-, medium-, and large-sized VSDs. Figure 9 (a) - (c) shows uncertainty in LV pressure-volume loops with each VSD size and across two years of age. We see that LV pressures decrease with VSD size at early age points. This is maintained throughout the rest of the two-year period. However, EDV and stroke volume are relatively higher at two years of age in large-sized VSDs (Figure 9 (c)), with larger pressure-volume loop area. In general, the uncertainties in cardiac dynamics are larger with early age, with relatively large uncertainty at later ages as well for medium- and large-sized VSDs. Simulations in the RV (Figure 9(d)-(f)) also show relatively larger uncertainty at early ages. Interestingly, the small-sized VSD simulations (Figure 9(d)) at one day of age have a nearly identical uncertainty to medium- and large-sized VSDs. The average RV pressure-volume loop shape at one day of age is distinct between VSD sizes, with a more triangular loop in the small-sized VSD case that becomes rounded for medium- and large-sized VSDs. There is a large reduction in pressure magnitudes and their uncertainty at five days of age, with less uncertainty as age increases. As expected, later time points suggest elevated RV pressure for the larger VSD sizes. We also see larger ESV, EDVs, and stroke volume for the large VSD simulations.

Figure 8. Variability in pulmonary vascular resistance ($R_p$, (a)), systemic vascular resistance ($R_s$, (b)), pulmonary vascular compliance ($C_{pa}$, (c)), and systemic vascular compliance ($C_{sa}$, (d)) when sampling maturation parameters in a female. Note that age on the x-axis is provided on a log-scale for better interpretability. Grey lines indicate 100 realizations while the solid black line and dashed red lines indicate the average and two standard deviations from the average, respectively.

We examined how VSD flow magnitudes (Figure 10) and the scalar metrics MAP, mPAP, and Qp/Qs (Figure 11) varied with uncertainty. VSD flow consistently shows higher relative uncertainty at early ages in comparison to later age points. All VSD sizes transition from having shunt flow in both directions to predominately left-to-right shunt flow (indicated by positive flow magnitudes) at five days of age. Large-sized VSDs still exhibit some bidirectional shunt flow at later ages, which is attributed to the uncertainty in $R_p$. Small- and medium-sized VSDs have relatively constant left-to-right shunt flow magnitudes with increasing age.

Figure 9. Variability in LV (a-c) and RV (d-f) pressure-volume loops for small (a,d), medium (b,e), and large (c,f) VSD sizes, corresponding to 1-, 3-, and 6-mm diameters in a female. Simulations use the pulmonary and systemic vascular parameter curves from Figure 8, as well as uncertain total blood volume. Grey lines indicate 100 realizations while the solid-colored lines indicate the average.

The scalar metrics provided in Figure 11 show the scaled probability density functions (PDFs) for MAP, mPAP, and Qp/Qs, where the PDF modes have been scaled to one for comparison. We see MAP values are nearly identical across VSD size at one day of age, with larger differences between the VSD sizes occurring at later time points. MAP values have much larger uncertainty compared to mPAP, which is attributed to the more gradual variability in $R_s$ and $C_{sa}$ as shown in Figure 8. The values of mPAP (Figure 11(b)) are nearly identical across

Figure 10. Variability in female VSD flow for small (a), medium (b), and large (c) VSD sizes, corresponding to 1-, 3-, and 6-mm diameters. Y-axes provide different magnitudes for ease of interpretability. Simulations use the pulmonary and systemic vascular parameter curves provided in Figure 8, as well as uncertain total blood volume. Grey lines indicate 100 realizations while the solid-colored lines indicate the average.

Figure 11. Variability in female mean systemic arterial pressure (MAP, (a)), mean pulmonary arterial pressure (mPAP, (b)), and the pulmonary-to-systemic flow ratio (Qp/Qs, (c)) as a function of age and VSD size. We use scaled-probability density functions, which are scaled such that the mode of each PDF is 1, for sake of interpretability. The width of the distribution indicates uncertainty in the estimate, and the y-axis provides the magnitude of the quantity. Simulations use the pulmonary and systemic vascular parameter curves and uncertain blood volumes from before.

VSD sizes at one day of age, but shift to distinct PDFs at five days of age. Larger VSDs have a rightward shift in their PDF reflecting increased mPAP values. The variance of mPAP is similar in all three VSD sizes, though large-sized VSDs do lead to slightly larger. The results for the Qp/Qs ratio (Figure 11(c)) illustrate more gradual changes in hemodynamics with age. The modes for all three PDFs overlap at 1 day of age, but the variance is larger for large-sized VSDs. This holds true through all age points, with the large-sized VSD having a variance nearly double that of the medium-sized VSD. As expected, small-sized VSDs have a Qp/Qs close to 1, whereas medium- and large-sized VSDs have Qp/Qs values in the ranges [1.2,1.5] and [1.8, 2.3], respectively.

## 4. Discussion

We simulated cardiovascular function across the first two years of life using an eight-compartment, lumped parameter model of the four heart chambers and major circulations. We employ a previously employed method for allometric and nonlinear scaling of model parameters with age to investigate the effects of VSD on hemodynamics. Larger defects increase flow across the VSD and pulmonary flow, increasing right-sided loading after the early neonatal period. Impaired PVR adaptation (modeled here as an arrested decline) sustains elevated RV pressures and higher mPAP. Uncertainty analysis on the scaling laws for PVR and SVR show substantial uncertainty in early age points, showing a significant impact of vascular resistance on early-age hemodynamics.

### 4.1 Simulating growth in a simple compartment model

Hiebing et al.[8] provided a foundational methodology for scaling PVR and SVR to simulate cardiovascular adaptation in newborns. A prior study by Munneke et al.[9] simulated the transition from fetal to neonatal circulations, including flow through the foramen ovale and ductus arteriosus, with a focus on oxygen saturation levels. Chen et al.[7] recently simulated hemodynamics in neonates with borderline LV, a condition where neonates are born with an underdeveloped LV that is on the border of being diagnosed as hypoplastic left heart syndrome. Chen et al. also used the adaptation strategy of Hiebing to scale model parameters for the circulation, and showed agreement with clinical data. The scaling laws for our simplistic eight-compartment model provide similar physiological outputs over two years of age, as shown in Figure 2. Predictions of MAP and mPAP are within the bounds of data collected in physiologically normal pediatric subjects across multiple measurement modalities [21–29]. Our results underperform in predicting ESVs, where we tend to have smaller heart values near the minimum bounds from prior data. However, the framework does capture the rapid drop in mPAP and a previously documented reduction in RV volumes immediately after birth with the drop in PVR[20].

Our adult simulations, used for scaling to pediatrics, are within the range of previously recording normotensive individuals[17,32–34]. We simulated sex-differences by providing slightly lower systemic and pulmonary pressures and smaller cardiac volumes in the female sex[20]. Our simulations of neonate hemodynamics are similar to adult hemodynamics in shape, but are drastically different in magnitude. Neonates have lower cardiac volumes, higher pulmonary pressures, and lower systemic pressures, consistent with our findings[9,23,24]. Neonatal pulmonary arterial pulse pressure is higher than the adult simulations due to lower $C_{pa}$ values, and simulated RV volumes are slightly larger than LV volumes. Our pediatric simulations of mitral valve flows show an E/a ratio, computed from the early and late flows into the LV, greater than 1, which increases with later ages. In contrast, the tricuspid E/a ratio is less than one in pediatrics, and then increases with later ages. This is consistent with prior echocardiography data[35] which showed that E/a increases with age and can be close to or less than one in pediatric subjects.

4.2 <u>Hemodynamic changes with VSD</u>

There are drastic changes in hemodynamic parameters at birth. These effects are modulated when a VSD is present[11,36]. The left-to-right shunt across the septum increases pulmonary vascular flows and changes hemodynamic loading[11,13]. Two key factors in determining left-to-right shunt are the VSD size and PVR, both contributing to the flow through the lesion. Our simulation results in Figure 4 assess the effects of VSD size at seven days of age, and show larger left-to-right flow magnitude with larger VSD sizes. Pressure-volume loops show increased EDV in both ventricles and the LA, with little change in RA volumes. Medium and large size VSDs often induce increased LA and LV preload and dilation[14,37], which is consistent with our simulations. The larger defect sizes increase LA pressure and LV stroke volume also. Our simulations also show a leftward shift in the RV, reflecting an increased EDV with increased RV pressures. While pressure increases are expected[37], increased RV EDV is uncommon. However, this can occur when PVR remains relatively low[37], which is consistent with our results since PVR drops quickly and is relatively low at seven days of age. Moreover, we do not explicitly include RV remodeling, and prior studies have shown that significant left-to-right shunts can lead to RV hypertrophy[36,38]. Our simulations show abnormal RV isovolumic contraction with a VSD, as RV volume increases during contraction due to the left-to-right shunt. Larger VSDs also increased the E/a value of the mitral valve, whereas the tricuspid valve flow is unaffected.

The interactions between PVR and VSD size dictate the hemodynamic state of a subject[7,29,37]. Figure 5 illustrated this interaction, as hemodynamic metrics at 30 days of age change drastically with VSD size and PVR values. The value of Qp/Qs is valuable in the clinical management of VSD[11,36,37], often used to determine the significance of VSD. For instance, larger Qp/Qs values have been associated with lower cerebral oxygen saturation[39]. We show that low to normal PVR can lead to a Qp/Qs>1.5 in medium and large-sized defects, but that higher PVR can reduce this ratio. We also quantify cardiac work, which provides insight into the effort needed by each cardiac chamber[40,41]. Given the difficulty of measuring pressure-volume loops in neonates, relatively few studies have documented these metrics. One study is Yang et al.[42], who compared clinical pressure-volume loops using pressure signals from right heart catheterization and volume estimates from magnetic resonance imaging in pediatric subjects with pulmonary hypertension. The authors showed that RV stroke work correlated with increased PVR over multiple time points in their patient population, similar to our findings. Works by Chowdhury et al.[43,44] used pressure-volume loops in the LV to assess contractility and diastolic stiffness in pediatric subjects, with results showing negligible differences in LV stroke work between those requiring and not requiring heart transplantation. Our results show that LV and RV stroke work move in opposite directions: larger VSDs and higher PVR reduce LV stroke work, while RV stroke work increases dramatically for large-sized VSDs and gradually with higher PVR. A large VSD with low or normal PVR has a high Qp/Qs, increasing RV volumes and stroke work. Though higher PVR decreases Qp/Qs, this elevates RV pressure and causes a steady increase in RV stroke work. These results suggest that the presence of a medium to large VSD drastically increases work done by the RV, and increased PVR due to limited pulmonary vascular adaptation would increase this workload. LA and RA stroke work are relatively novel and rarely used metrics, but may provide insight into cardiac function in this subject group[45,46]. The use of stroke work in general for assessing neonate function is uncommon, but may provide insights beyond current diagnostics[41,42].

4.3 <u>Effects of VSD across age</u>

Detection and diagnosis of VSD by echocardiography is critical at the early stages of life. A computational model can provide significant additional insight if it can simulate hemodynamic trajectories with adaptation at later ages. In the case of medium-sized VSDs, the decision to operate can be riddled with uncertainty[11,12,16], providing an avenue for assistance by computational modeling. The LV pressure-volume loops in Figure 6(a) show the natural changes in hemodynamics with and without a VSD. Both LV EDV and ESV increase with age, indicative of natural weight gain and increased blood volumes, along with increased contractility. LV pressures naturally rise with increased SVR and reduced systemic vascular compliance, consistent with prior findings[9,47]. In the presence of a medium- or large-sized VSD, we observe larger LV EDV, with a prominent increase in LV stroke volume at the later age points. This increase in stroke volume coincides with reduced LV pressure and highlights the model's ability to reproduce LV volume overload with larger VSDs[11,12]. The RV undergoes a rapid transition during the initial stages of life in the absence of a VSD, with significantly larger systolic pressures at one day of age, followed by a drop in pressure magnitudes and eventual increase in blood volumes without the presence of a VSD. Small-sized VSDs cause little to no change in this trajectory. However, medium- and large-sized VSDs cause a delayed decrease in RV pressures, in addition to a rightward shift indicative of higher EDVs. The RV continues to have higher ESPs at the later age points, suggesting that a large-sized VSD in the absence of elevated PVR can still lead to pulmonary hypertensive conditions. Larger RV EDVs suggest dilation of the chamber, which is not a hallmark of untreated VSDs, in contrast to LV and LA dilation[36,38]. We attribute this discrepancy in part to our lack of explicit mechanisms for cardiovascular remodeling, which has been considered in several other studies focusing on cardiac[8] or vascular[30,48] adaptation. However, we found that doubling $E_{min}$ in RV, reflecting an increased stiffness with hypertrophy, or decreasing systemic venous compliance, $C_{SV}$, by 25% to increase volume to the systemic circulation reduced RV volume overload in our simulations. This result, provided in the Supplementary Material (Figure S1), supports the role of cardiac and vascular adaptation in the progression of untreated VSDs and should be investigated further.

Pulmonary vascular adaptation is a major factor in VSD management. We considered stunted adaptation, where pulmonary arterial parameters were fixed at their 30-day or three-day values. The former, shown in Figure 7(a)-(b), leads to increased stroke volume in the LV with all VSD cases as was shown in Figure 6. For the RV, we see biphasic behavior. From one day to 30 days of age, we see a reduction in RV ESP and a mild increase in RV EDV for the large-sized VSDs. Then, during stunted adaptation, we see an increase in RV ESP at one year of age in all scenarios, with an even larger increase in RV pressures and EDV at two years of age for the large-sized VSD. A similar trend is seen in Figure 7(c)-(d) when pulmonary vascular parameters adapt until three days of age. The LV undergoes a smaller increase in EDV while the RV shows a significantly higher ESP at 30 days of age. This holds for the non-VSD and all VSD sizes. The biphasic fluctuation in RV pressure which may occur in subjects with congenital heart defects with stunted pulmonary vascular adaptation [49–51] may lead to delayed recognition of pulmonary vascular disease in these subjects.

4.4 <u>Effects of pulmonary and systemic arterial parameters</u>

Given that the adaptation curves from Hiebing et al. were constructed based on data from multiple studies, we investigated how uncertainties in the growth parameters contribute to differences in hemodynamics for our model framework. This is similar to a sensitivity analysis, which eight-compartment models have been subject to in prior studies[17,52,53]. Most studies, though, focus on a single time point without adaptation. One exception is the study by Jones and Oomen[54], which considered a compartment level model with biventricular interaction and cardiac growth and remodeling. The authors calibrated and validated cardiac remodeling trajectories, under uncertainty, to data from an animal model of mitral valve regurgitation. Otherwise, uncertainty in adaptation rules have had limited investigation from a sensitivity or uncertainty analysis perspective.

Like Jones and Oomen, we sampled parameters describing adaptation and propagated these uncertainties over multiple time points. This provides us with more of a "population" level view of our simulated hemodynamics. As shown in Figure 8, uncertainty in the vascular resistance and compliance parameters is much larger at younger ages. However, we do not see a drastic reduction in uncertainty in our simulated outputs, as shown in Figures 9-11. In particular, the LV pressures in Figure 9 have similar uncertainty across all age points, while LV volume uncertainty appears larger at later age points. This is attributed to our inclusion of total blood volume as a random variable and reflects the natural variability in blood volumes due to weight gain and heterogeneous development in children. Interestingly, we see that LV pressure uncertainty in medium- and large-sized VSDs (Figure 9(b)-(c)) is relatively smaller than that in small-sized VSDs (Figure 9(a)), while the converse is true about volume variability. In contrast to the LV, pressure variability in the RV is much larger at one day of age than at later time points, though there is still substantial fluctuations in volumes at two years of age. The interactions between VSD size, hemodynamic parameters, and total blood volume contribute to the severity of VSD-affected congenital heart defects[37]. The larger uncertainty at early age points coincides with our understanding of VSD diagnosis: the size of the VSD and subject-specific parameters of the neonate (e.g., elevated PVR and low weight gain[14,16]) interact to determine the hemodynamic loading of the heart. Our study also shows that both LV and RV volumes can vary significantly at later age points with the presence of a medium- or large-sized VSD. Volume overload is a risk factor in the development of heart failure in children with non-restrictive VSDs[14,16], consistent with our predictions of elevated cardiac volumes in Figure 9.

Flow across the VSD, shown in Figure 10, again supports the claim that earlier age points have relatively higher uncertainty. Flow across a VSD can be quantified by echocardiogram, but is relatively hard to measure exactly[11,36,37]. In general, the larger the defect, the greater the magnitude of the left-to-right shunt. At the initial age point, we see nearly equal magnitude flows from left-to-right and from right-to-left. This coincides with a larger PVR and similar magnitudes of LV and RV pressure, thus allowing flow in both directions. This quickly dissipates at five days of age as PVR begins to fall. Our systems-level results in Figure 11 show that systemic hemodynamics are relatively consistent across age points and VSD size. As expected, higher pulmonary arterial pressures track with large-sized VSDs. The uncertainty in these quantities tend to decrease with age as well. The Qp/Qs values also vary with age and VSD size. As we saw in Figure 10, the Qp/Qs values are relatively similar in their average value but have larger uncertainty for larger VSD sizes. At five days of age, when PVR has dropped, we achieved a distinct shift in the PDFs for Qp/Qs, with medium- and large-sized VSDs being centered around 1.5 and 2, respectively. These results are again consistent with guidelines on managing VSD, with Qp/Qs values in the range of 1.6 – 2.8 serving as clinical cutoffs[11,55]. Thus, our results show that medium and large-sized defects reproduce Qp/Qs ratios that are consistent with those seen in the clinical management of VSD.

4.5 <u>Limitations</u>

There are multiple limitations for the present study. First, this initial study provides evidence that relatively simple hemodynamic models, combined with adaptation laws and VSD flow, can reproduce hemodynamics with similar magnitudes to measured data. We did not explicitly calibrate or validate our model to data obtained across VSD subjects, which is the next necessary step in transitioning our approach to a digital shadow and, eventually, a digital twin[56]. We consider a simplified model of the four heart chambers, though future modeling studies should incorporate biventricular interaction[9] to identify how septal wall motion is augmented with the defect. We have previously used sensitivity analysis on a similar compartmental model to quantify predictive uncertainty in the context of adult pulmonary hypertension[17]. However, the present model has nested levels of parameter interactions: there are parameters that affect hemodynamics at a single age, and parameters that ultimately dictate how resistance, compliance, and elastance change over time. Future studies that develop multiscale global sensitivity methods that explicitly account for temporal correlations are warranted. We imposed potential vascular pathologies by elevating PVR and reducing pulmonary artery compliance, though vascular remodeling should emerge because of hemodynamics. Incorporating growth and remodeling into the framework explicitly is a necessary next step[8]. We found that left-to-right shunts volume overloaded the RV, which would be corrected with a proper growth and remodeling framework. Finally, we only considered lumped parameter representations of the systemic and pulmonary arteries and veins. Including additional compartments, such as the brain, liver, or coronary circulation[7], would increase the complexity of the problem but may help identify specific relationships between VSD size and compartment-specific hemodynamics[39].

## 5. Conclusion

This study provides a computational platform for examining the effects of small-, medium-, and large-sized VSDs on hemodynamics from one day to two years old. We used a relatively simplistic model with a minimal number of parameters to identify how age-dependent parameterizations, combined with various VSD size, alter hemodynamic loading in the cardiac chambers. Our results are consistent with prior measurements in neonates and provide an initial study for using computational modeling to understand VSD management. This is a necessary step towards the use of computer models in subject-specific management of isolated VSD and other congenital heart diseases which coincide with VSDs in neonates.

## 6. Acknowledgements

We would like to thank Ashley A. Hiebing and Colleen Witzenburg for providing data from their original study.

## 7. Data Accessibility

The code for this work can be found at https://github.com/mjcolebank/VSD_Simulation.

## References

1.  Ko BS, Cameron JD, Munnur RK, et al. Noninvasive CT-Derived FFR Based on Structural and Fluid Analysis. *JACC Cardiovasc Imaging*. 2017;10(6):663-673. doi:10.1016/j.jcmg.2016.07.005

2. Schiavazzi DE, Kung EO, Marsden AL, et al. Hemodynamic effects of left pulmonary artery stenosis after superior cavopulmonary connection: A patient-specific multiscale modeling study. *Journal of Thoracic and Cardiovascular Surgery*. 2015;149(3):689-696.e3. doi:10.1016/j.jtcvs.2014.12.040

3. Schiavazzi DE, Arbia G, Baker C, et al. Uncertainty quantification in virtual surgery hemodynamics predictions for single ventricle palliation. *Int J Numer Method Biomed Eng*. 2016;32(3):e02737. doi:10.1002/cnm.2737

4. Taylor-LaPole AM, Colebank MJ, Weigand JD, Olufsen MS, Puelz C. A computational study of aortic reconstruction in single ventricle patients. *Biomech Model Mechanobiol*. 2023;22(1):357-377. doi:10.1007/s10237-022-01650-w

5. Taylor-LaPole AM, Paun LM, Lior D, Weigand JD, Puelz C, Olufsen MS. Parameter selection and optimization of a computational network model of blood flow in single-ventricle patients. *J R Soc Interface*. 2025;22(223). doi:10.1098/rsif.2024.0663

6. Schiavazzi DE, Baretta A, Pennati G, Hsia TY, Marsden AL. Patient-specific parameter estimation in single-ventricle lumped circulation models under uncertainty. *Int J Numer Method Biomed Eng*. 2017;33(3):1-34. doi:10.1002/cnm.2799

7. Chen Y, Anzai IA, Kalfa DM, Vedula V. A Patient-Specific Computational Model for Neonates and Infants with Borderline Left Ventricles. *Ann Biomed Eng*. 2026;54(1):41-61. doi:10.1007/s10439-025-03894-w

8. Hiebing AA, Pieper RG, Witzenburg CM. A Computational Model of Ventricular Dimensions and Hemodynamics in Growing Infants. *J Biomech Eng*. 2023;145(10). doi:10.1115/1.4062779

9. Munneke AG, Lumens J, Delhaas T. Cardiovascular fetal-to-neonatal transition: an in silico model. *Pediatr Res*. 2022;91(1):116-128. doi:10.1038/s41390-021-01401-0

10. Shimizu S, Une D, Kawada T, et al. Lumped parameter model for hemodynamic simulation of congenital heart diseases. *Journal of Physiological Sciences.Springer Tokyo*. 2018;68(2):103-111. doi:10.1007/s12576-017-0585-1

11. Eleyan L, Zaidi M, Ashry A, Dhannapuneni R, Harky A. Ventricular septal defect: Diagnosis and treatments in the neonates: A systematic review. *Cardiol Young*. 2021;31(5):756-761. doi:10.1017/S1047951120004576

12. Schipper M, Slieker MG, Schoof PH, Breur JMPJ. Surgical Repair of Ventricular Septal Defect; Contemporary Results and Risk Factors for a Complicated Course. *Pediatr Cardiol*. 2017;38(2):264-270. doi:10.1007/s00246-016-1508-2

13. Valente AM, Freed MD. Half a century of experience with congenital heart disease: what have we learned about ventricular septal defects? *Eur Heart J.Oxford University Press*. 2023;44(1):62-63. doi:10.1093/eurheartj/ehac657

14. Safa R, Dean A, Sanil Y, Thomas R, Singh G, Charaf Eddine A. Effect of Preoperative Volume Overload on Left Ventricular Function Recovery After Ventricular Septal Defect Repair. *American Journal of Cardiology*. 2023;203:253-258. doi:10.1016/j.amjcard.2023.06.118

15. Eckerström F, Nyboe C, Redington A, Hjortdal VE. Lifetime Burden of Morbidity in Patients With Isolated Congenital Ventricular Septal Defect. *J Am Heart Assoc*. 2023;12(1). doi:10.1161/JAHA.122.027477

16. Eckerström F, Nyboe C, Maagaard M, Redington A, Hjortdal VE. Survival of patients with congenital ventricular septal defect. *Eur Heart J*. 2023;44(1):54-61. doi:10.1093/eurheartj/ehac618

17. Colunga AL, Colebank MJ, Olufsen MS. Parameter inference in a computational model of haemodynamics in pulmonary hypertension. *J R Soc Interface*. 2023;20(200). doi:10.1098/rsif.2022.0735

18. Olufsen M, Nadim A. On deriving lumped models for blood flow and pressure in the systemic arteries. *Mathematical Biosciences and Engineering*. 2004;1(1):61-80. doi:10.3934/mbe.2004.1.61

19. Pahuja M, Schrage B, Westermann D, Basir MB, Garan AR, Burkhoff D. Hemodynamic Effects of Mechanical Circulatory Support Devices in Ventricular Septal Defect: Results from a Computer Simulation Model. *Circ Heart Fail*. 2019;12(7). doi:10.1161/CIRCHEARTFAILURE.119.005981

20. Boron WF, Boulpaep EL. *Medical Physiology*. Philadelphia, PA : Elsevier, [2017]; 2017. http://catalog.lib.ncsu.edu/catalog/NCSU3738615

21. Thilenius OG, Arcilla RA. Angiographic Right and Left Ventricular Volume Determination in Normal Infants and Children. *Pediatr Res*. 1974;8(2):67-74. doi:10.1203/00006450-197402000-00001

22. GRAHAM TP, JARMAKANI JM, ATWOOD GF, CANENT R V. Right Ventricular Volume Determinations in Children. *Circulation*. 1973;47(1):144-153. doi:10.1161/01.CIR.47.1.144

23. Kent AL, Kecskes Z, Shadbolt B, Falk MC. Blood pressure in the first year of life in healthy infants born at term. *Pediatric Nephrology*. 2007;22(10):1743-1749. doi:10.1007/s00467-007-0561-8

24. Kent AL, Kecskes Z, Shadbolt B, Falk MC. Normative blood pressure data in the early neonatal period. *Pediatric Nephrology*. 2007;22(9):1335-1341. doi:10.1007/s00467-007-0480-8

25. Lytrivi ID, Bhatla P, Ko HH, et al. Normal values for left ventricular volume in infants and young children by the echocardiographic subxiphoid five-sixth area by length (bullet) method. *Journal of the American Society of Echocardiography*. 2011;24(2):214-218. doi:10.1016/j.echo.2010.12.002

26. Kang C, Zhao E, Zhou Y, et al. Dynamic changes of pulmonary arterial pressure and ductus arteriosus in human newborns from birth to 72 Hours of Age. *Medicine (United States)*. 2016;95(3). doi:10.1097/MD.0000000000002599

27. Qi HY, Ma RY, Jiang LX, et al. Anatomical and hemodynamic evaluations of the heart and pulmonary arterial pressure in healthy children residing at high altitude in China. *IJC Heart and Vasculature*. 2015;7:158-164. doi:10.1016/j.ijcha.2014.10.015

28. Olivieri LJ, Jiang J, Hamann K, et al. Normal right and left ventricular volumes prospectively obtained from cardiovascular magnetic resonance in awake, healthy, 0- 12 year old children. *Journal of Cardiovascular Magnetic Resonance*. 2020;22(1):11. doi:10.1186/S12968-020-0602-Z

29. Emmanouilides GC, Moss AJ, Duffie ER, Adams FH. Pulmonary arterial pressure changes in human newborn infants from birth to 3 days of age. *J Pediatr*. 1964;65(3):327-333. doi:10.1016/S0022-3476(64)80395-4

30. Dong ML, Lan IS, Yang W, Rabinovitch M, Feinstein JA, Marsden AL. Computational simulation-derived hemodynamic and biomechanical properties of the pulmonary arterial tree early in the course of ventricular septal defects. *Biomech Model Mechanobiol*. 2021;20(6):2471-2489. doi:10.1007/s10237-021-01519-4

31. Bambul Heck P, Eicken A, Kasnar-Samprec J, Ewert P, Hager A. Early pulmonary arterial hypertension immediately after closure of a ventricular or complete atrioventricular septal defect beyond 6 months of age. *Int J Cardiol*. 2017;228:313-318. doi:10.1016/j.ijcard.2016.11.056

32. Colunga AL, Kim KG, Woodall NP, et al. Deep phenotyping of cardiac function in heart transplant patients using cardiovascular system models. *J Physiol*. 2020;598(15):3203-3222. doi:10.1113/JP279393

33. Lumens J, Delhaas T, Kirn B, Arts T. Three-Wall Segment (TriSeg) Model Describing Mechanics and Hemodynamics of Ventricular Interaction. *Ann Biomed Eng*. 2009;37(11):2234-2255. doi:10.1007/s10439-009-9774-2

34. Kim SM, Randall EB, Jezek F, Beard DA, Chesler NC. Computational modeling of ventricular-ventricular interactions suggest a role in clinical conditions involving heart failure. *Front Physiol*. 2023;14. doi:10.3389/fphys.2023.1231688

35. Heusinkveld MHG, Huberts W, Lumens J, Arts T, Delhaas T, Reesink KD. Large vessels as a tree of transmission lines incorporated in the CircAdapt whole-heart model: A computational tool to examine heart-vessel interaction. Marsden A, ed. *PLoS Comput Biol*. 2019;15(7):e1007173. doi:10.1371/journal.pcbi.1007173

36. Cantinotti M, Marchese P, Scalese M, et al. Normal Values and Patterns of Normality and Physiological Variability of Mitral and Tricuspid Inflow Pulsed Doppler in Healthy Children. *Healthcare (Switzerland)*. 2022;10(2). doi:10.3390/healthcare10020355

37. Spicer DE, Hsu HH, Co-Vu J, Anderson RH, Fricker FJ. Ventricular septal defect. *Orphanet J Rare Dis*. 2014;9(1):144. doi:10.1186/s13023-014-0144-2

38. Amer SM, El Nemr SB, El Amrosy DM, Elrasol OAR. Speckle tracking echocardiography of right and left ventricles in children of atrial septal defect and ventricular septal defect. *Tanta Medical Journal*. 2024;52(4):352-356. doi:10.4103/tmj.tmj_35_24

39. Banno A, Kanazawa T, Shimizu K, et al. Higher preoperative Qp/Qs ratio is associated with lower preoperative regional cerebral oxygen saturation in children with ventricular septal defect. *J Anesth*. 2021;35(3):442-445. doi:10.1007/s00540-021-02931-x

40. Menting ME, Cuypers JAAE, Opić P, et al. The Unnatural History of the Ventricular Septal Defect. *J Am Coll Cardiol*. 2015;65(18):1941-1951. doi:10.1016/j.jacc.2015.02.055

41. Driessen MMP, Leiner T, Sieswerda GT, et al. RV adaptation to increased afterload in congenital heart disease and pulmonary hypertension. *PLoS One*. 2018;13(10):1-18. doi:10.1371/journal.pone.0205196

42. Kheyfets V, Thirugnanasambandam M, Rios L, et al. The role of wall shear stress in the assessment of right ventricle hydraulic workload. *Pulm Circ*. 2015;5(1):90-100. doi:10.1086/679703

43. Schramm W. The units of measurement of the ventricular stroke work: a review study. *J Clin Monit Comput*. 2010;24(3):213-217. doi:10.1007/s10877-010-9234-4

44. Yang W, Marsden AL, Ogawa MT, et al. Right ventricular stroke work correlates with outcomes in pediatric pulmonary arterial hypertension. *Pulm Circ*. 2018;8(3):204589401878053. doi:10.1177/2045894018780534

45. Chowdhury SM, Butts RJ, Hlavacek AM, et al. Echocardiographic Detection of Increased Ventricular Diastolic Stiffness in Pediatric Heart Transplant Recipients: A Pilot Study. *Journal of the American Society of Echocardiography*. 2018;31(3):342-348.e1. doi:10.1016/j.echo.2017.11.010

46. Chowdhury SM, Butts RJ, Taylor CL, et al. Longitudinal measures of deformation are associated with a composite measure of contractility derived from pressure-volume loop analysis in children. *Eur Heart J Cardiovasc Imaging*. 2018;19(5):562-568. doi:10.1093/ehjci/jex167

47. Geelhoed JJM, Steegers EAP, Osch-Gevers L van, et al. Cardiac structures track during the first 2 years of life and are associated with fetal growth and hemodynamics. The Generation R Study. *Am Heart J*. 2009;158(1):71-77. doi:10.1016/j.ahj.2009.04.018

48. Yang W, Dong M, Rabinovitch M, Chan FP, Marsden AL, Feinstein JA. Evolution of hemodynamic forces in the pulmonary tree with progressively worsening pulmonary arterial hypertension in pediatric patients. *Biomech Model Mechanobiol*. 2019;18(3):779-796. doi:10.1007/s10237-018-01114-0

49. Jone PN, Ivy DD, Hauck A, et al. Pulmonary Hypertension in Congenital Heart Disease: A Scientific Statement From the American Heart Association. *Circ Heart Fail*. 2023;16(7). doi:10.1161/HHF.0000000000000080

50. Pascall E, Tulloh RMR. Pulmonary hypertension in congenital heart disease. *Future Cardiol.Future Medicine Ltd.* 2018;14(3):369-375. doi:10.2217/fca-2017-0065

51. Lammers AE, Bauer LJ, Diller GP, et al. Pulmonary hypertension after shunt closure in patients with simple congenital heart defects. *Int J Cardiol*. 2020;308:28-32. doi:10.1016/j.ijcard.2019.12.070

52. Colebank MJ, Chesler NC. An in-silico analysis of experimental designs to study ventricular function: A focus on the right ventricle. Marsden AL, ed. *PLoS Comput Biol*. 2022;18(9):e1010017. doi:10.1371/journal.pcbi.1010017

53. Colebank MJ, Taylor R, Hacker TA, Chesler NC. Biventricular Interaction During Acute Left Ventricular Ischemia in Mice: A Combined In-Vivo and In-Silico Approach. *Ann Biomed Eng*. 2023;51(11):2528-2543. doi:10.1007/s10439-023-03293-z

54. Jones CE, Oomen PJA. Synergistic biophysics and machine learning modeling to rapidly predict cardiac growth probability. *Comput Biol Med*. 2025;184. doi:10.1016/j.compbiomed.2024.109323

55. National Academies of Sciences E and Medicine. *Foundational Research Gaps and Future Directions for Digital Twins*. National Academies Press; 2024. doi:10.17226/26894

**Table 1.** Nominal estimates of blood volumes and blood pressures for the model for both sex. Nominal adult hemodynamic parameters are derived from formulae provided.

| Parameter | Description | Units | Nominal Value or Equation |
|---|---|---|---|
| **Nominal Volumes** (Adult Female/Male) | | | |
| $V_{LA}^0$ | Initial LA volume | mL | 20/25 |
| $V_{LV}^0$ | Initial LV volume | mL | 100/120 |
| $V_{SA}^0$ | Initial SA volume | mL | 182.25/202.5 |
| $V_{SV}^0$ | Initial SV volume | mL | 675/750 |
| $V_{RA}^0$ | Initial RA volume | mL | 15/25 |
| $V_{RV}^0$ | Initial RV volume | mL | 100/120 |
| $V_{PA}^0$ | Initial PA volume | mL | 234.9/261 |
| $V_{PV}^0$ | Initial PV volume | mL | 49.5/55 |
| **Nominal Pressures** (Adult Female/Male) | | | |
| $p_{SA,max}$ | Maximum SA pressure | mmHg | 110/120 |
| $p_{SA,min}$ | Minimum SA pressure | mmHg | 70/80 |
| $\bar{p}_{SA}$ | Mean SA pressure | mmHg | 83.33/93.33 |
| $p_{PA,max}$ | Maximum PA pressure | mmHg | 17/20 |
| $p_{PA,min}$ | Minimum PA pressure | mmHg | 5/8 |
| $\bar{p}_{PA}$ | Mean PA pressure | mmHg | 12/13 |
| $\bar{p}_{SV}$ | Mean SV pressure | mmHg | 8/10 |
| $\bar{p}_{PV}$ | Mean PV pressure | mmHg | 4/4 |
| $p_{LA,max}$ | Maximum LA pressure | mmHg | 13/16 |
| $p_{LA,min}$ | Minimum LA pressure | mmHg | 3/3 |
| $p_{LV,max}$ | Maximum LV pressure | mmHg | 111/121 |
| $p_{LV,min}$ | Minimum LV pressure | mmHg | 5/5 |
| $p_{RA,max}$ | Maximum RA pressure | mmHg | 6/8 |
| $p_{RA,min}$ | Minimum RA pressure | mmHg | 2/2 |
| $p_{RV,max}$ | Maximum RV pressure | mmHg | 18/21 |
| $p_{RV,min}$ | Minimum RV pressure | mmHg | 3/3 |
| **Vascular System Parameters** | | | |
| $R_{a,val}$ | Aortic valve resistance | mmHg s / mL | $0.1 \div CO$ |
| $R_{m,val}$ | Mitral valve resistance | mmHg s / mL | $0.5 \div CO$ |
| $R_{p,val}$ | Pulmonic valve resistance | mmHg s / mL | $0.1 \div CO$ |
| $R_{t,val}$ | Tricuspid valve resistance | mmHg s / mL | $0.5 \div CO$ |
| $R_{vc}$ | Vena Cava resistance | mmHg s / mL | $(\bar{p}_{SV} - p_{RA,min}) \div CO$ |
| $R_{pv}$ | Pulmonary venous | mmHg s / mL | $(\bar{p}_{PV} - p_{LA,min}) \div CO$ |

| | | resistance | | |
|---|---|---|---|---|
| $R_s$ | Systemic circulation resistance | | mmHg s / mL | $(\bar{p}_{SA} - \bar{p}_{SV,}) \div CO$ |
| $R_p$ | Pulmonary circulation resistance | | mmHg s / mL | $(\bar{p}_{PA} - \bar{p}_{PV}) \div CO$ |
| $R_{VSD}$ | Ventricular septal defect resistance | | mmHg s / mL | $\left(\pi r_{VSD}^2 \cdot \sqrt{p^* \cdot 2/\rho}\right)^{-1}$ |
| $C_{sa}$ | Compliance of systemic arteries | | mL / mmHg | $SV \div (p_{SA,max} - p_{SA,min})$ |
| $C_{sv}$ | Compliance of systemic veins | | mL / mmHg | $V_{sv} \div \bar{p}_{sv}$ |
| $C_{pa}$ | Compliance of pulmonary arteries | | mL / mmHg | $SV \div (p_{PA,max} - p_{PA,min})$ |
| $C_{pv}$ | Compliance of pulmonary veins | | mL / mmHg | $V_{pv} \div \bar{p}_{pv}$ |
| **Cardiac Parameters** | | | | |
| $E_{max,LA}$ | Maximum LA Elastance | | mmHg / mL | $p_{LA,max} \div V_{LA,min}$ |
| $E_{min,LA}$ | Minimum LA Elastance | | mmHg / mL | $(p_{LA,max} + p_{LA,max}) \div (V_{LA,min} + V_{LA,max})$ |
| $E_{max,LV}$ | Maximum LV Elastance | | mmHg / mL | $p_{LV,max} \div V_{LV,min}$ |
| $E_{min,LV}$ | Minimum LV Elastance | | mmHg / mL | $p_{LV,min} \div V_{LV,max}$ |
| $E_{max,RA}$ | Maximum RA Elastance | | mmHg / mL | $p_{RA,max} \div V_{RA,min}$ |
| $E_{min,RA}$ | Minimum RA Elastance | | mmHg / mL | $(p_{RA,max} + p_{RA,max}) \div (V_{RA,min} + V_{RA,max})$ |
| $E_{max,RV}$ | Maximum RV Elastance | | mmHg / mL | $p_{RV,max} \div V_{RV,min}$ |
| $E_{min,RV}$ | Minimum RV Elastance | | mmHg / mL | $p_{RV,min} \div V_{RV,max}$ |
| $T_{max,LA}$ | Time of maximum LA elastance | | s | $0.1 \cdot T$ |
| $T_{min,LA}$ | Time of minimum LA elastance | | s | $0.2 \cdot T$ |
| $T_{max,LV}$ | Time of maximum LV elastance | | s | $0.25 \cdot T$ |
| $T_{min,LV}$ | Time of minimum LV elastance | | s | $0.55 \cdot T$ |
| $T_{max,RA}$ | Time of maximum RA | | s | $0.1 \cdot T$ |

|  | elastance |  |  |
|---|---|---|---|
| $T_{min,RA}$ | Time of minimum RA elastance | s | $0.2 \cdot T$ |
| $T_{max,RV}$ | Time of maximum RV elastance | s | $0.25 \cdot T$ |
| $T_{min,RV}$ | Time of minimum RV elastance | s | $0.55 \cdot T$ |
| $T_{off}$ | Offset time for atrial contraction | s | $0.15 \cdot T$ |

**Table 2.** Allometric- and maturation-based growth parameters. Those calculated explicitly in this paper was denoted by "Determined."

| Parameter | Description | Units | Value (Ref) |
|---|---|---|---|
| **Allometric Scaling** | | | |
| $b_{BV}$ | Blood volume | ND | 1.0 (Hiebing[8]) |
| $b_{HR}$ | Heart rate | ND | −0.25 (Hiebing[8]) |
| $b_{E_{max}}$ | Maximum elastance ($E_{max}$) | ND | −1.0 (Hiebing[8]) |
| $b_{E_{min}}$ | Minimum elastance ($E_{min}$) | ND | −0.5 (Determined) |
| $b_{T_{max}}$ | Maximum ventricular elastance ($T_{max}$) | ND | −0.07 (Hiebing[8]) |
| $b_R$ | Vascular resistance ($R_p, R_{vc}, R_{pv}$) | ND | −0.75 (Hiebing[8]) |
| $b_{R_s}$ | Systemic vascular resistance ($R_p$) | ND | −0.5 (Determined) |
| $b_{C_v}$ | Venous compliance ($C_{sv}$ and $C_{pv}$) | ND | 1.0 (Hiebing[8]) |
| $b_{C_A}$ | Arterial compliance ($C_{sa}$ and $C_{pa}$) | ND | −1.33 (Hiebing[8]) |
| **Maturation Parameters** (Systemic and Pulmonary Resistance) | | | |
| $g_{s,1}$ | Early-age systemic vascular resistance | mmHg s / mL | 3.8 (Determined) |
| $g_{s,2}$ | Systemic vascular resistance time constant | days | 50 (Determined) |
| $g_{s,3}$ | Systemic vascular resistance scaling term | ND | 1.1 (Determined) |
| $g_{p,1}$ | Early-age pulmonary vascular resistance | mmHg s / mL | 4.0 (Determined) |
| $g_{p,2}$ | Pulmonary vascular resistance time constant | days | 1.0 (Determined) |
| $g_{p,3}$ | Pulmonary vascular resistance scaling term | ND | 2.0 (Determined) |

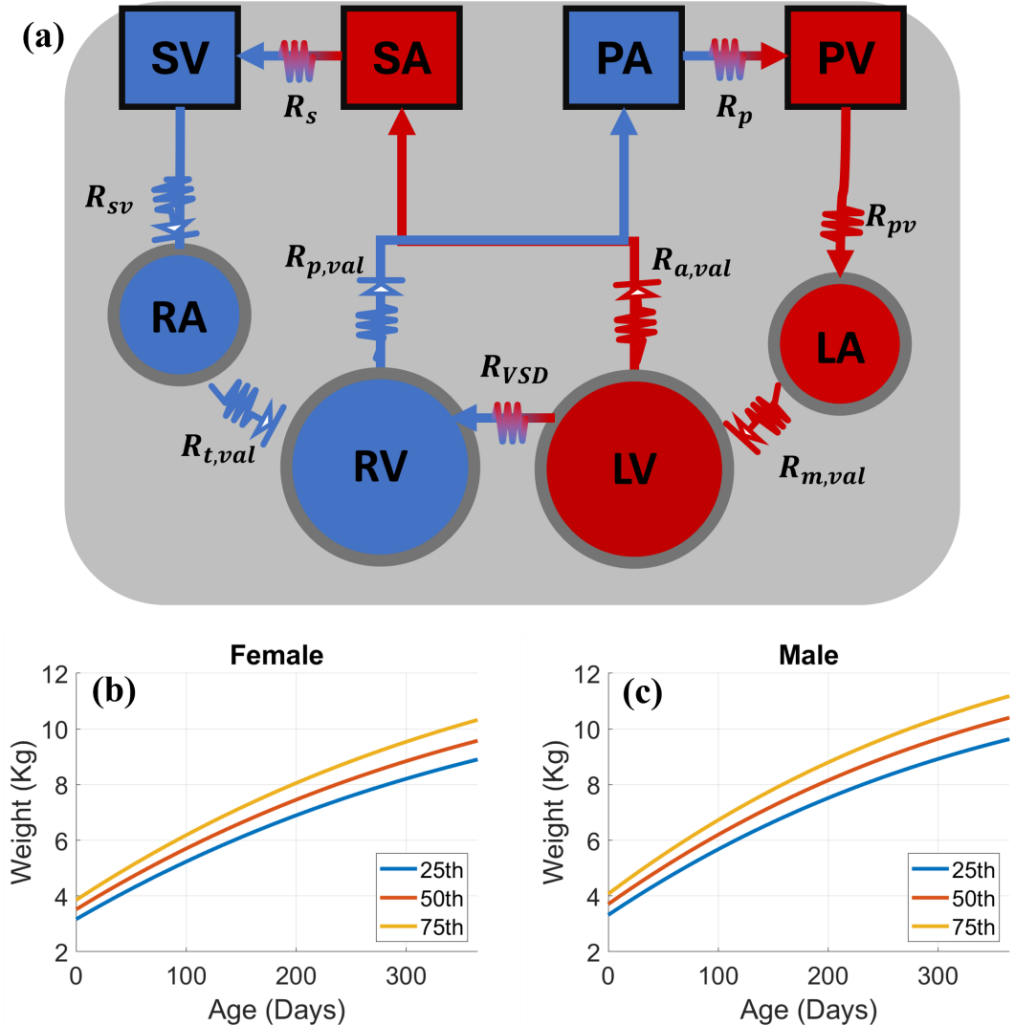

**Figure 1.** (a) Schematic of the mathematical model. The left atrial (LA), left ventricle (LV), right atrium (RA), and right ventricle (RV) are described by elastance functions, and connected to the systemic and pulmonary arteries (SA, PA) and veins (SV, PV) through resistors. The circulation is modeled as a compliance chamber. (b-c) Growth trajectories for female and male individuals (respectively) through one year of life based on CDC data.

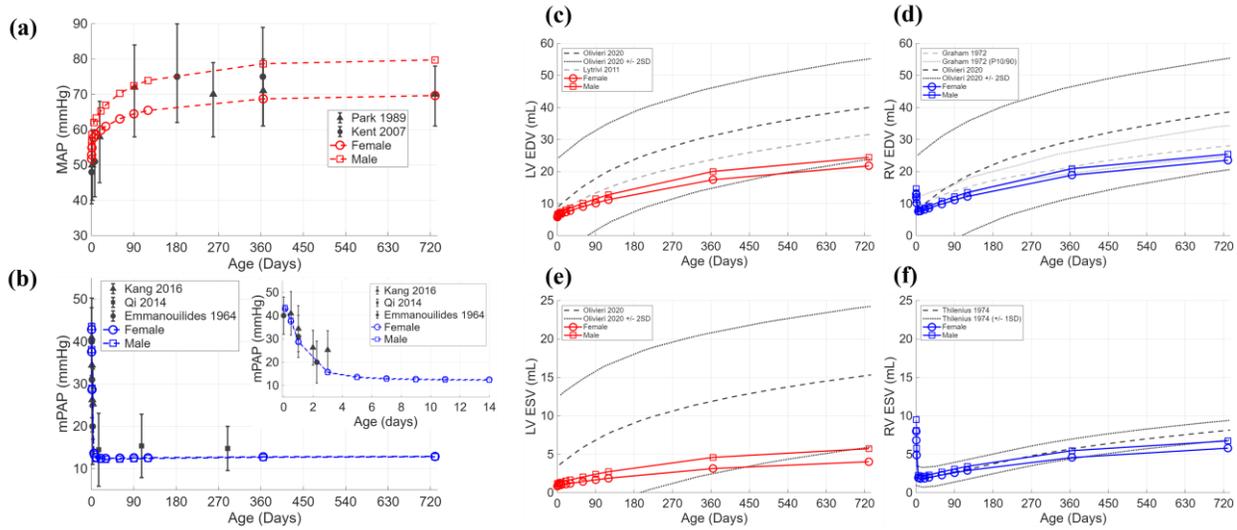

**Figure 2.** Changes in (a) MAP and (b) mPAP over the first two years of life in both females and males. Data from multiple studies[21–29] reproduced in the study by Heiberg are also presented. A subplot of mPAP during the first two weeks of life is also provided in (b). End-diastolic volumes (EDV, (c) and (d)) and end-systolic volumes (ESV, (e) and (f)) are presented over the same time-frame as subplots (a) and (b). Data from multiple studies[21–29] reproduced in the study by Hiebing are also presented.

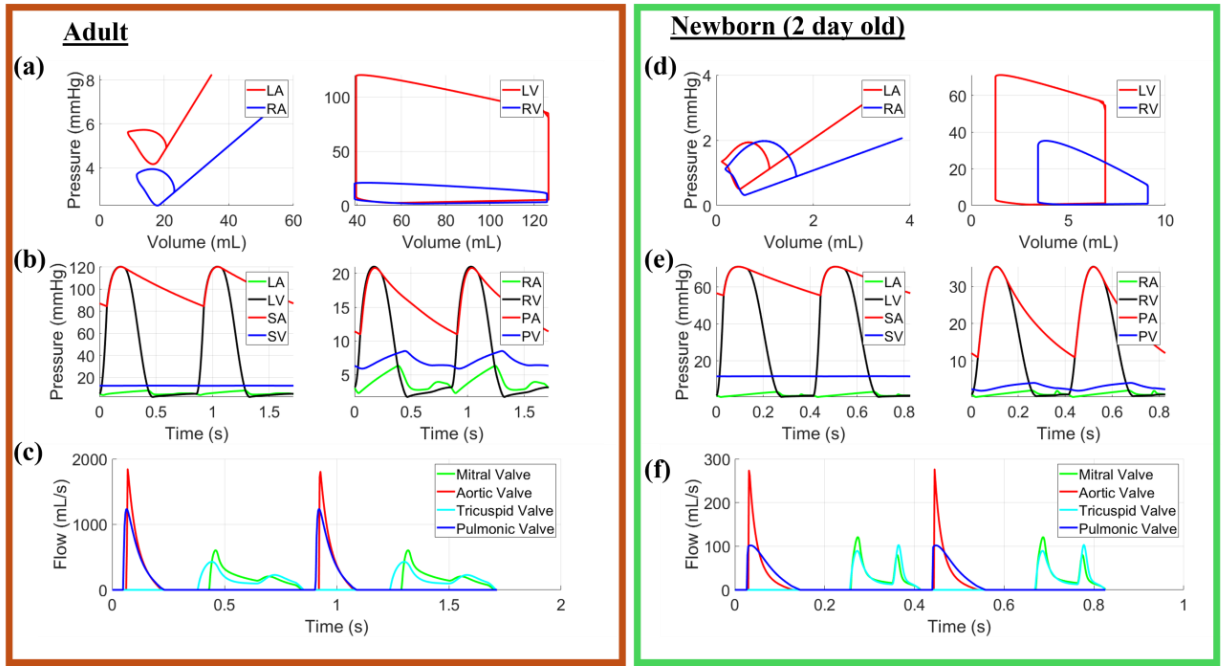

**Figure 3.** (a-c) Simulations in a typical adult, 70 kg Male, including atrial and ventricular pressure-volume loops (a), vascular pressures (b), and valvular flows (c). (d-f) Simulations from the same model but with parameterizations based on allometric and maturation-based scaling laws to a two-day old. Note that pulmonary pressures are significantly higher, with smaller cardiac volumes. See Figure 1 for abbreviations for the model infrastructure.

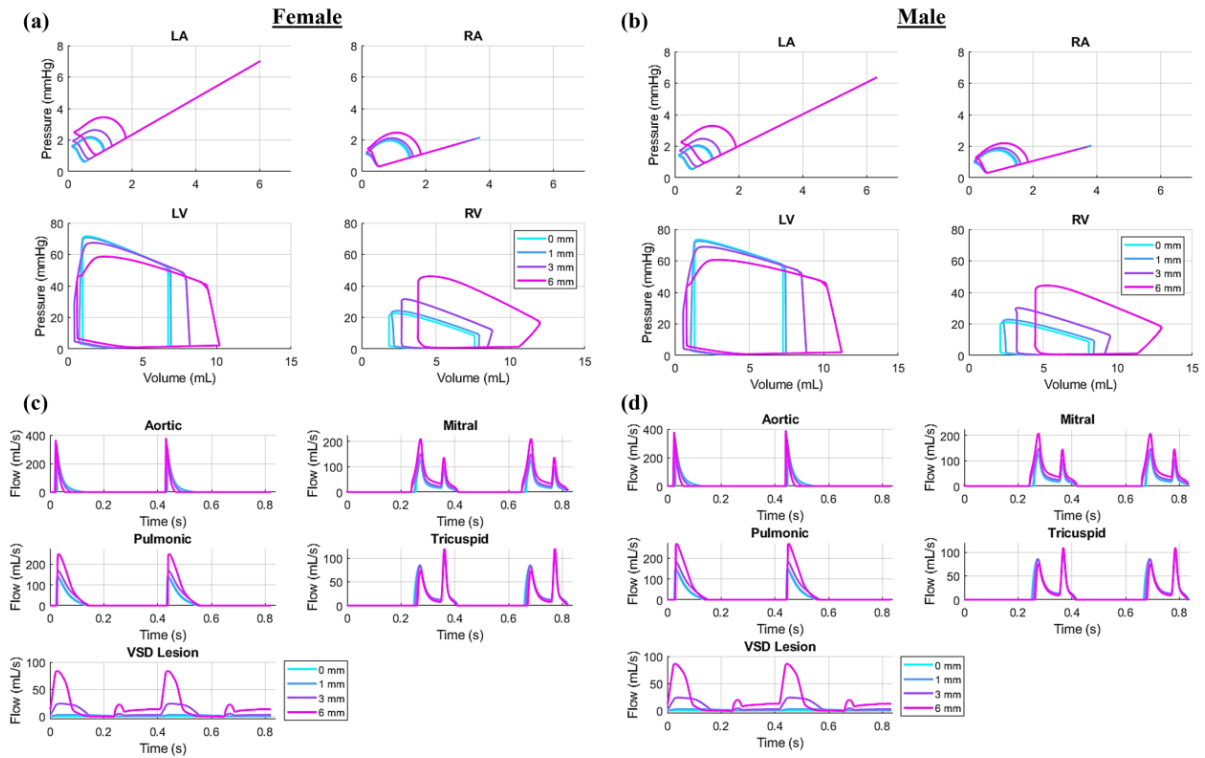

**Figure 4.** Simulations in a seven-day old pediatric female (a,c) and male (b,d) subject with and without a VSD. Cardiac function (a-b) and valvular flows (c-d) are shown for each sex with no VSD, a small VSD, a medium VSD, and a large VSD, corresponding to 0, 1, 3, and 6 mm in diameter, respectively. Subplots (c) and (d) include flow across the VSD, with positive flow indicated left-to-right shunting.

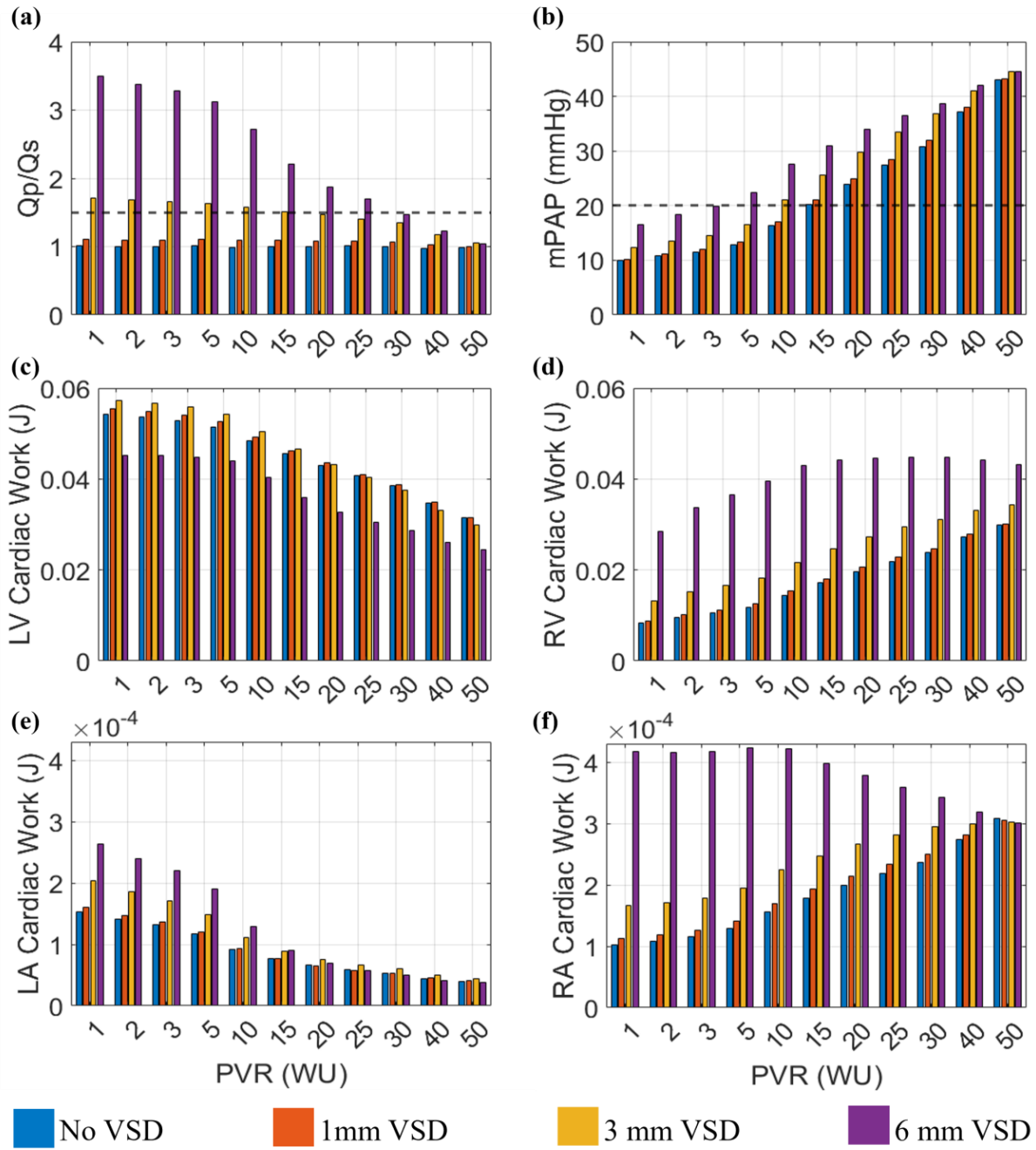

**Figure 5.** Metrics of hemodynamic and cardiac function for various VSD size (provided as diameter) and PVR magnitudes, presented in Wood Units (WU), for a 30 day old male subject. (a) Ratio of mean pulmonary flow to mean systemic flow (Qp/Qs), with the threshold of 1.5 provided. (b) Mean pulmonary arterial pressure (mPAP), with the threshold of 20 mmHg used to define pulmonary hypertension. (c-f) LV, RV, LA, and RA cardiac work calculated from simulated pressure-volume loops.

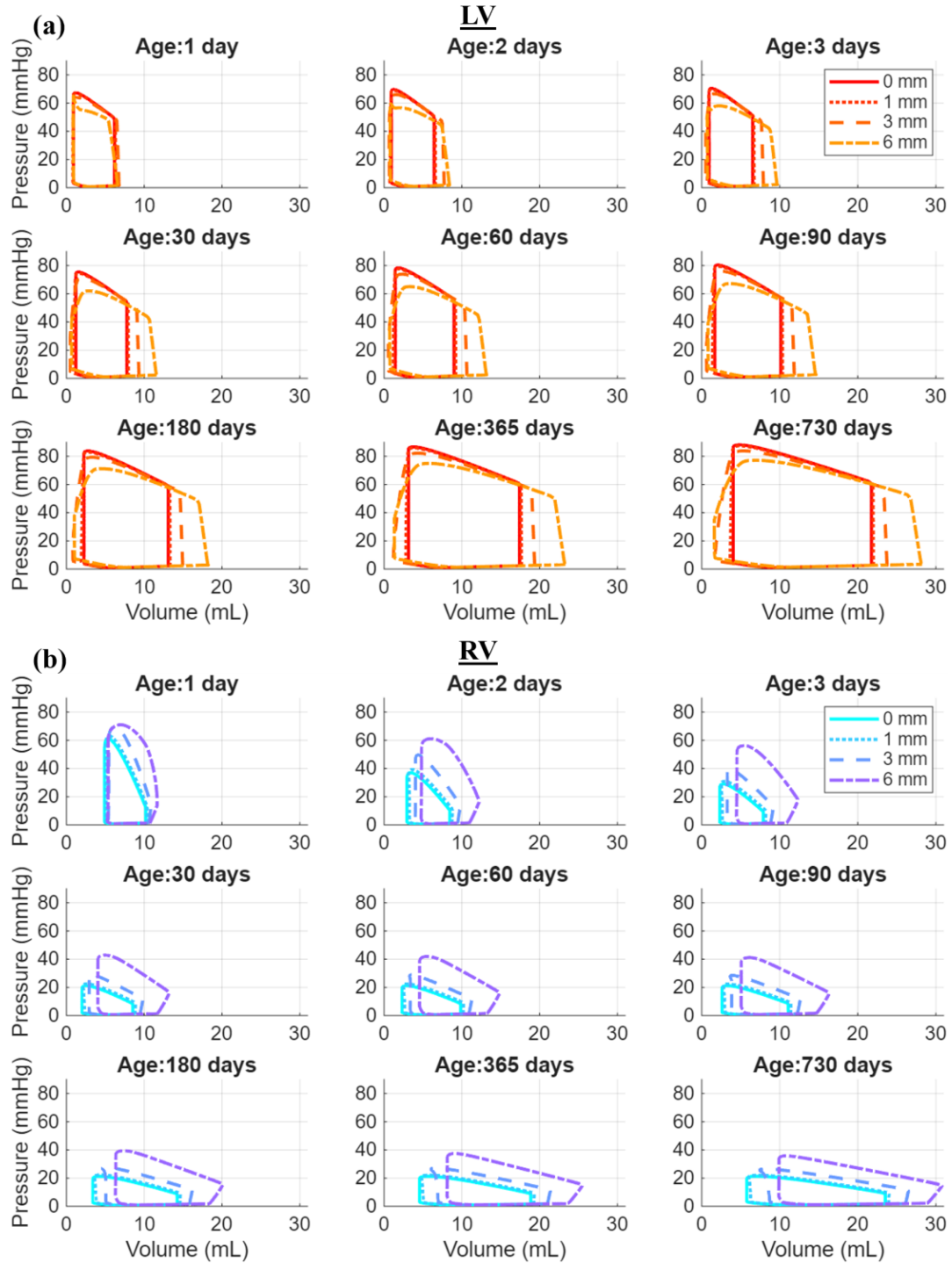

**Figure 6.** Ventricular pressure-volume loops as a function of age and VSD severity in a female. Parameters of the model are scaled using allometric and maturation-based approaches as described in the text. (a) Hemodynamics in the LV from one day to two years of age. VSD diameters used in the simulations are 0, 1, 3, and 6 mm in diameter. (b) Hemodynamics in the RV over the same time span.

**Figure 7.** Ventricular pressure-volume loops as a function of age and VSD severity with different pulmonary vascular conditions in a female. All parameters of the model are scaled as described in the text except for $R_p$ and $C_{pa}$. (a-b) Hemodynamics in the LV and RV from 1 day to two years of age under the assumption that $R_p$ and $C_{pa}$ change until 30 days of age. (b) Corresponding RV predictions. (c-d) Hemodynamics in the LV and RV from 1 day to two years of age under the assumption that $R_p$ and $C_{pa}$ are fixed at the value obtained at 3 days of age.

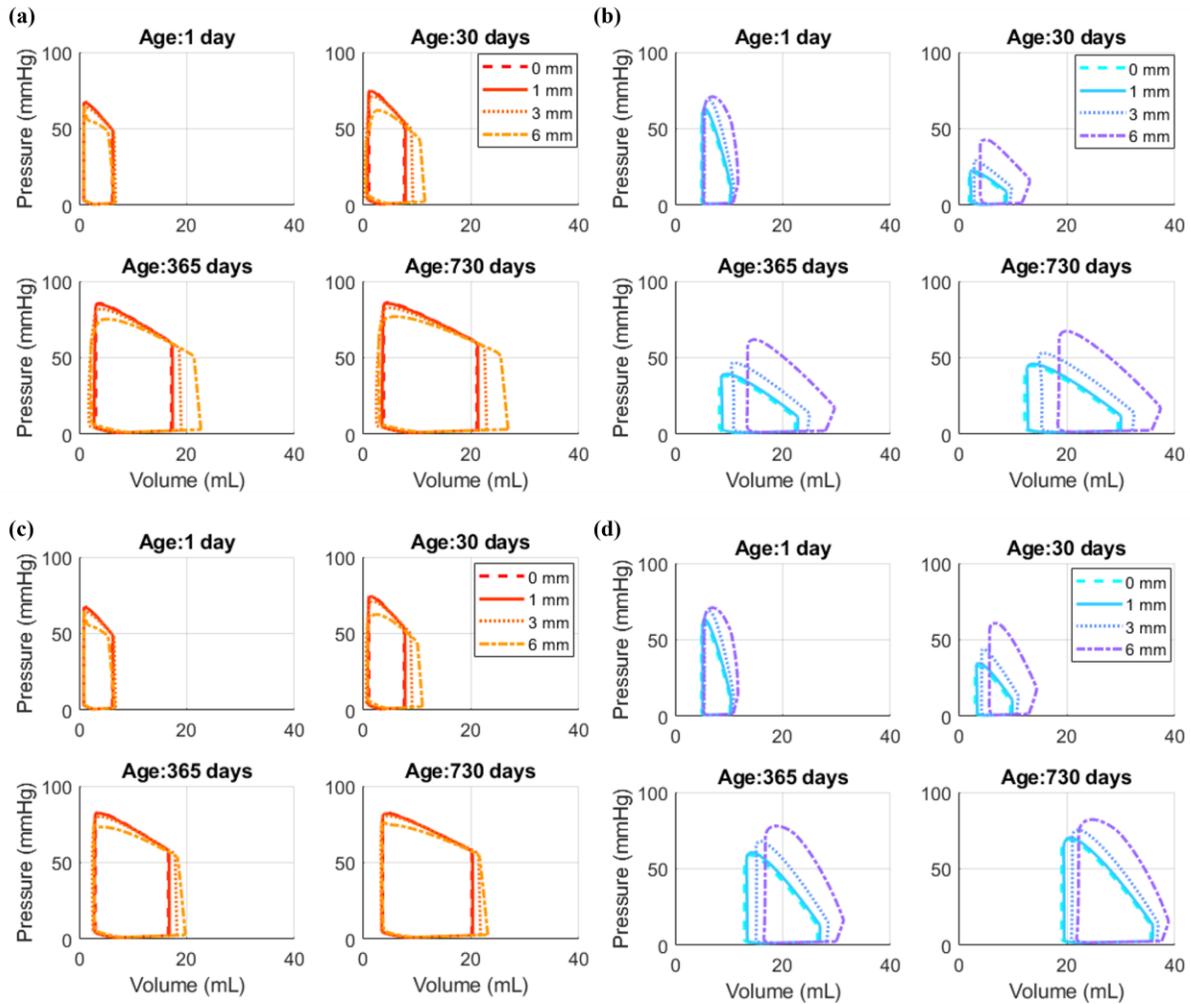

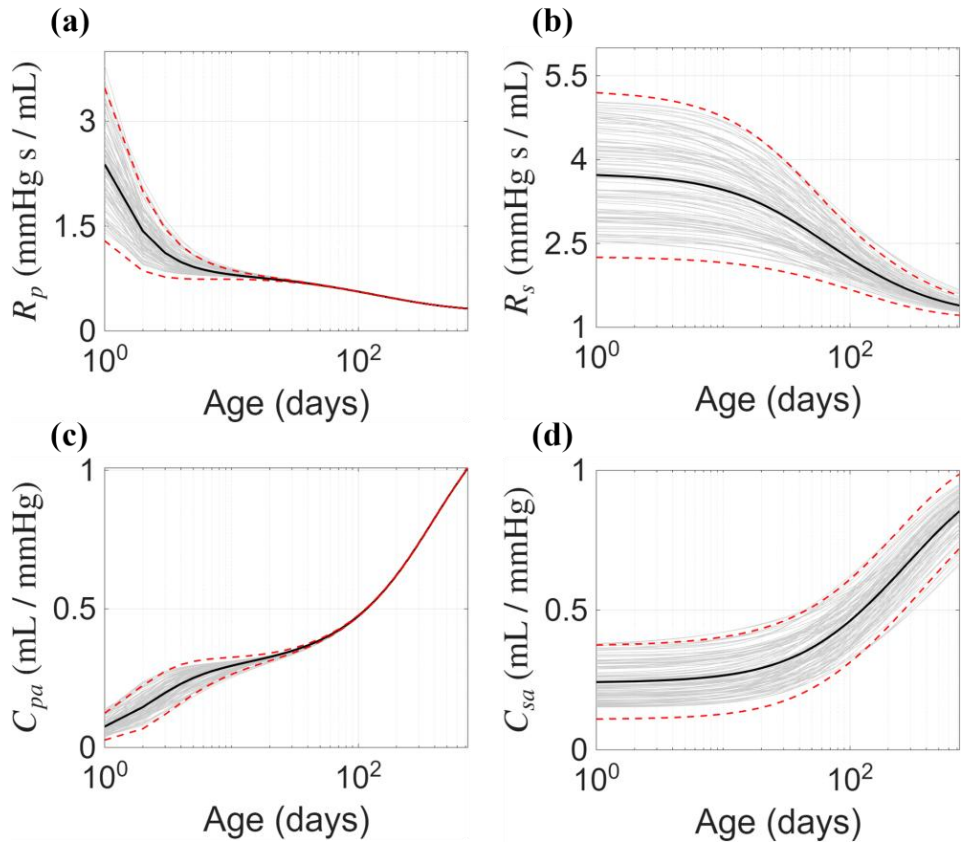

**Figure 8.** Variability in pulmonary vascular resistance ($R_p$, (a)), systemic vascular resistance ($R_s$, (b)), pulmonary vascular compliance ($C_{pa}$, (c)), and systemic vascular compliance ($C_{sa}$, (d)) when sampling maturation parameters in a female. Note that age on the x-axis is provided on a log-scale for better interpretability. Grey lines indicate 100 realizations while the solid black line and dashed red lines indicate the average and two standard deviations from the average, respectively.

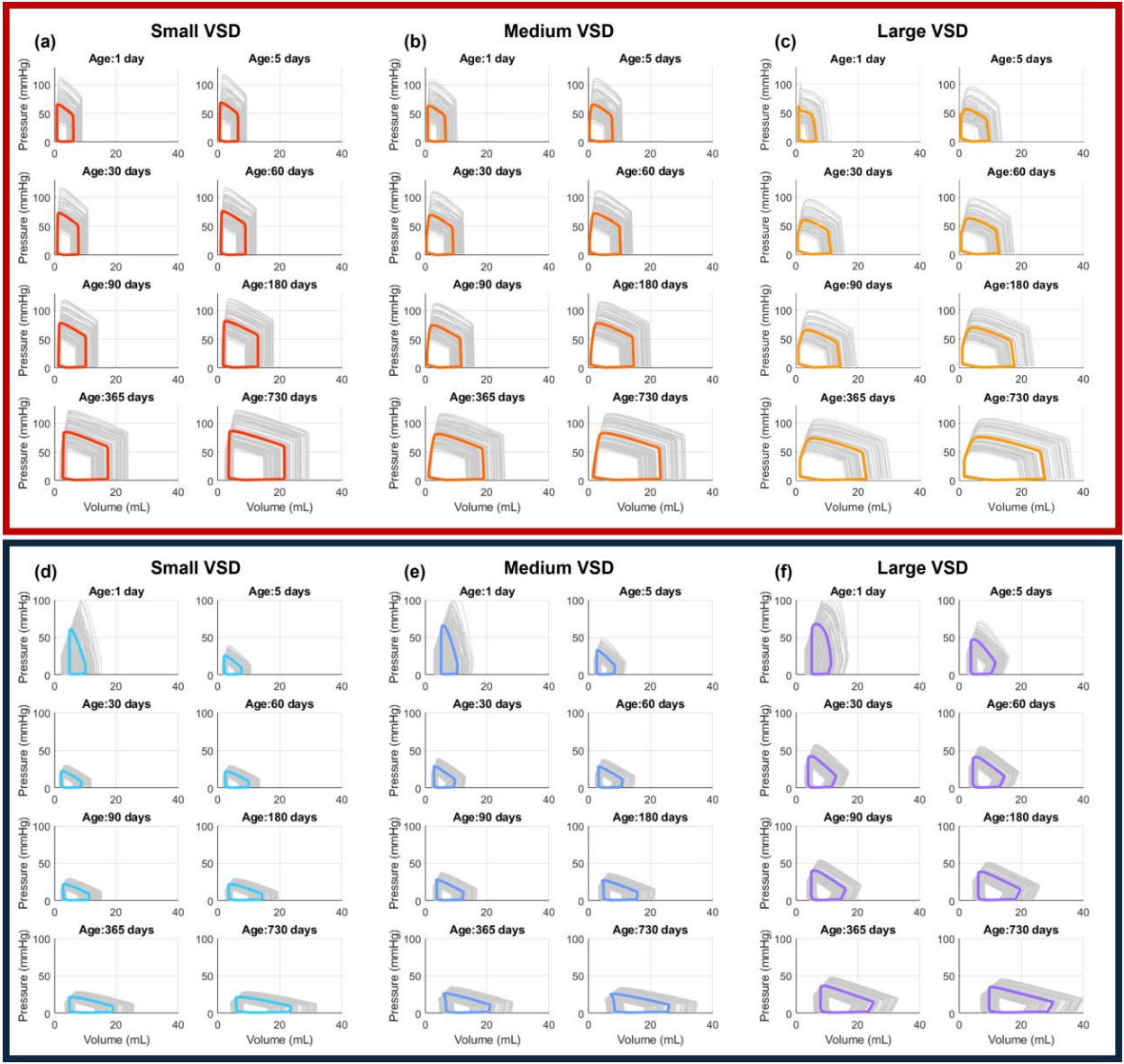

**Figure 9.** Variability in LV (a-c) and RV (d-f) pressure-volume loops for small (a,d), medium (b,e), and large (c,f) VSD sizes, corresponding to 1-, 3-, and 6-mm diameters in a female. Simulations use the pulmonary and systemic vascular parameter curves from Figure 8, as well as uncertain total blood volume. Grey lines indicate 100 realizations while the solid-colored lines indicate the average.

**Figure 10.** Variability in female VSD flow for small (a), medium (b), and large (c) VSD sizes, corresponding to 1-, 3-, and 6-mm diameters. Y-axes provide different magnitudes for ease of interpretability. Simulations use the pulmonary and systemic vascular parameter curves provided in Figure 8, as well as uncertain total blood volume. Grey lines indicate 100 realizations while the solid-colored lines indicate the average.

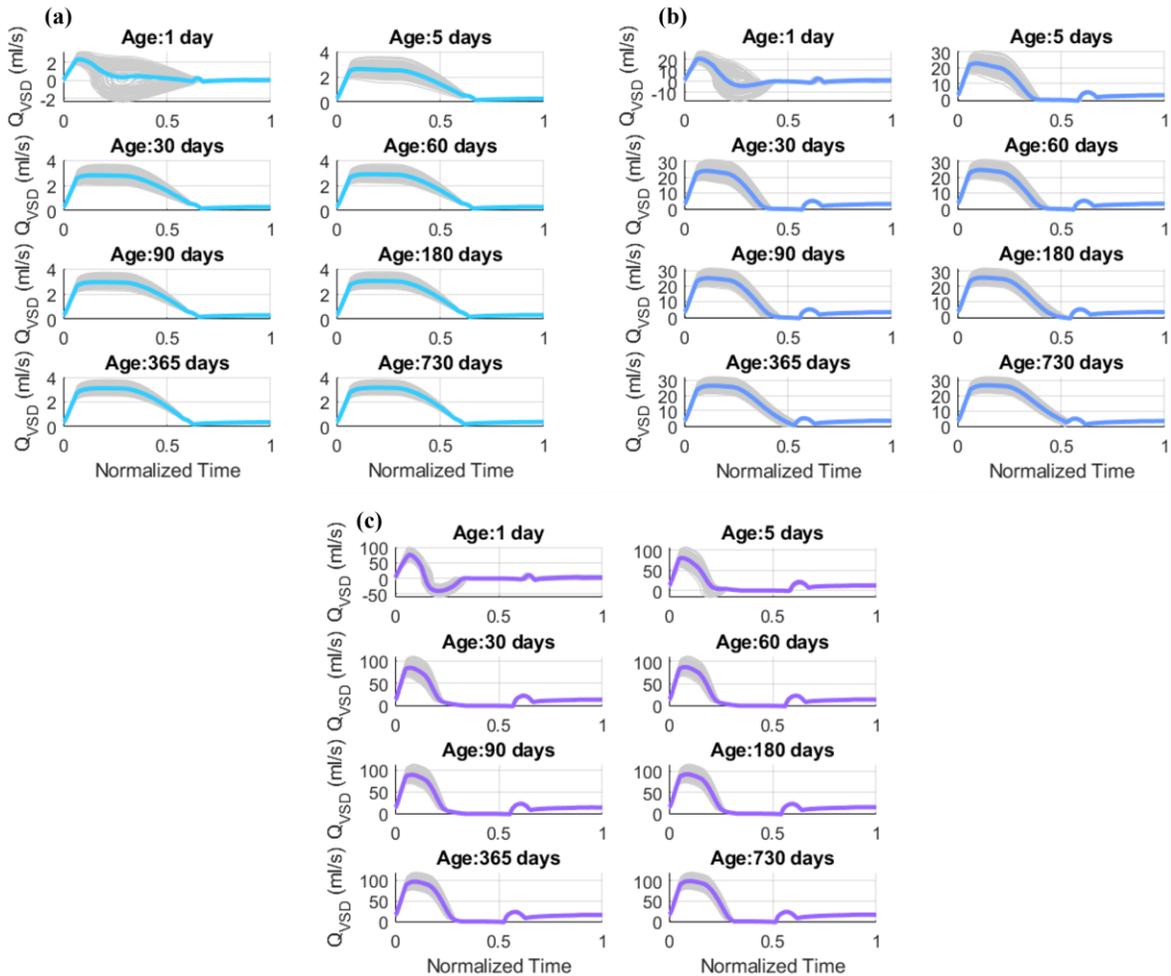

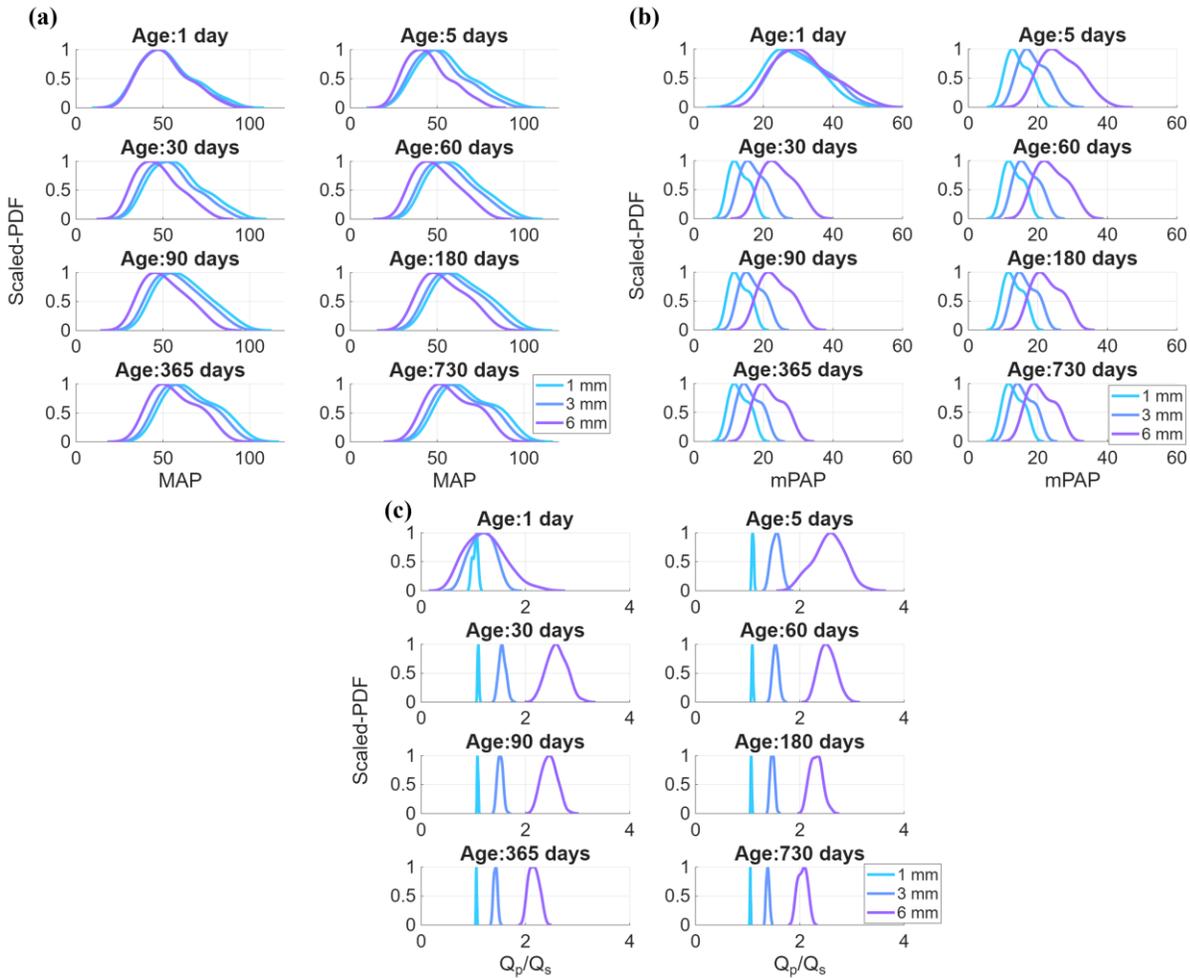

**Figure 11.** Variability in female mean systemic arterial pressure (MAP, (a)), mean pulmonary arterial pressure (mPAP, (b)), and the pulmonary-to-systemic flow ratio (Qp/Qs, (c)) as a function of age and VSD size. We use scaled-probability density functions, which are scaled such that the mode of each PDF is 1, for sake of interpretability. The width of the distribution indicates uncertainty in the estimate, and the y-axis provides the magnitude of the quantity. Simulations use the pulmonary and systemic vascular parameter curves and uncertain blood volumes from before.

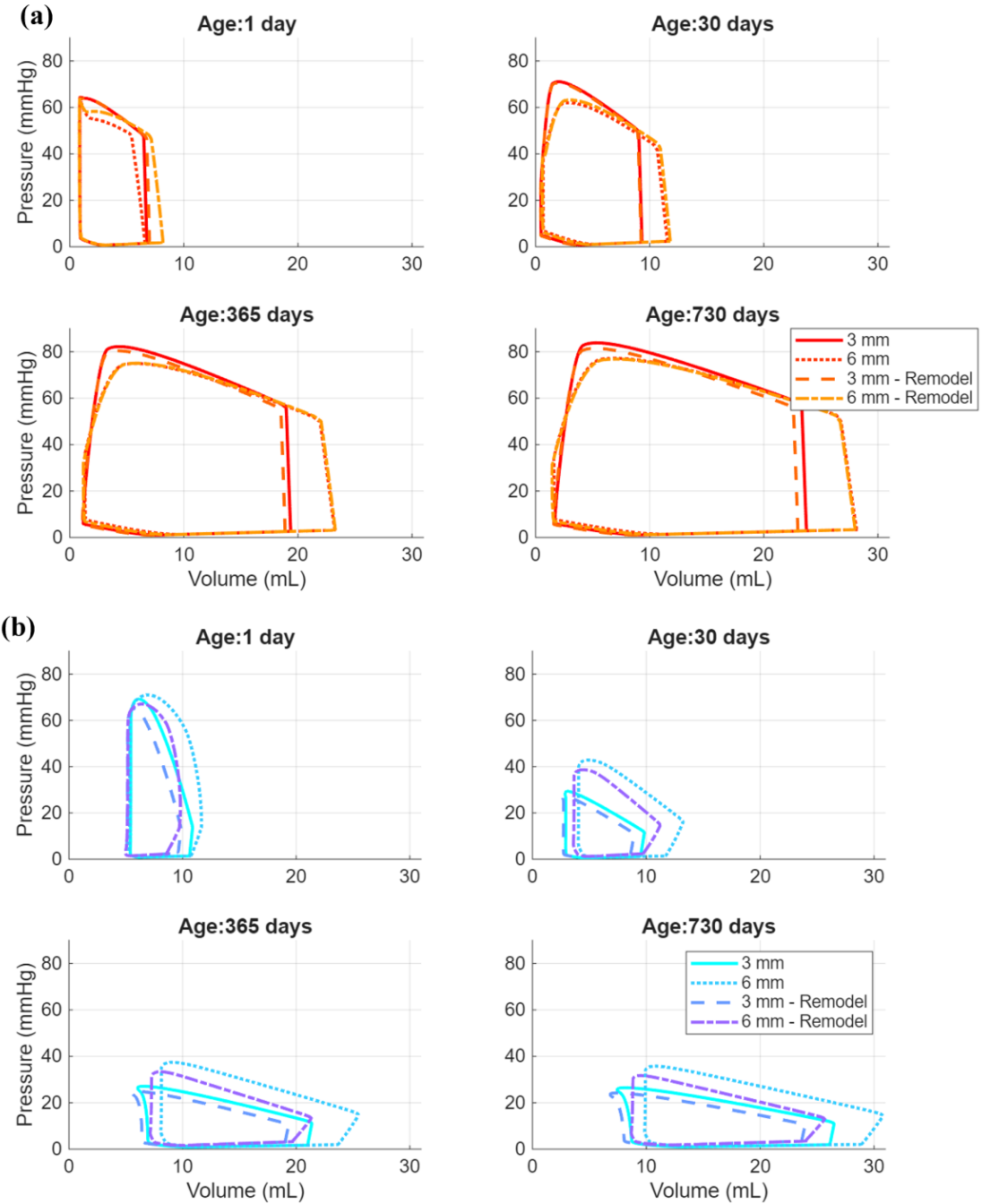

**Figure S1.** Effects of RV remodeling on hemodynamic predictions in the LV (a) and RV (b) of a female, at one-, 30-, 365-, and 730-days of age. The medium- and large-sized VSDs are simulated without RV adaptation, and then with a doubling of minimum elastance, $E_{min}$, in the RV to reflect stiffening of the ventricle.